\documentclass[11pt,a4paper]{article}
\usepackage{epsf}
\makeatletter \@addtoreset{equation}{section} \makeatother

\begin{document}

\def\be{\begin{equation}} 
\def\ee{\end{equation}}
\def\TMP#1#2#3{{\sl Theor. Mat. Phys.} {\bf #1} (#2) #3}
\def\PRB#1#2#3{{\sl Phys. Rev.} {\bf B#1} (#2) #3} 
\def\NPB#1#2#3{{\sl Nucl. Phys.} {\bf B#1} (#2) #3} 
\def\IJMPB#1#2#3{{\sl Int. J. Mod. Phys.} {\bf B#1} (#2) #3}
\def\mb{\bar{\mu}} 
\def\da{\downarrow}
\def\up{\uparrow} 
\def\be{\begin{equation}} 
\def\ee{\end{equation}}
\def\bea{\begin{eqnarray}} 
\def\eea{\end{eqnarray}}
\def\eps{\varepsilon} 
\def\g{\gamma} 
\def\b{\beta} 
\def\bp{{b^\prime}}
\def\d{\delta}
\def\a{\alpha} 
\def\ap{{a^\prime}} 
\def\e{\varepsilon} 
\def\l{\lambda}
\def\lp{\lambda^\prime} 
\def\s{\sigma} 
\def\La{\Lambda}
\def\nn{\nonumber\\} 
\def\gt{\widetilde{G}} 
\def\G{\Gamma}
\def\w{\langle W|} 
\def\tA{\tilde A} 
\def\tB{\tilde B}
\def\vr{\varrho} 
\def\v{|V\rangle} 
\def\wt{\langle {\tilde W}|}
\def\vt{|{\tilde V}\rangle} 
\def\r#1{(\ref{#1})}
\def\sm{{\bar Q}} 
\def\cq{{\cal Q}} 
\def\t#1{\langle\tau_{#1}\rangle}
\def\2t#1#2{\langle\tau_{#1}\tau_{#2}\rangle} 
\def\up{\uparrow}
\def\1l{\lambda^{(1)}}
\def\emph#1{{\sl #1}}
\begin{titlepage}
\def\thefootnote{\fnsymbol{footnote}}
\begin{flushright}
\setlength{\baselineskip}{13pt}
ITP-UH-22/96 \hfill September 1996 \\		
OUTP-9633S \hfill 
cond-mat/9610222
\end{flushright}
\vspace*{\fill}

\begin{center}
{\Large\sc Exact solution of a $t$-$J$ chain with impurity}\\
      
\vfill
\vspace{1.5 em} 
{\sc Gerald Bed\"urftig}$^1$ \footnote{e-mail: {\tt bed@itp.uni-hannover.de}}, 
{\sc Fabian H.L. E\char'31ler}$^2$\footnote{e-mail: {\tt fab@thphys.ox.ac.uk}}
{\sc and Holger Frahm}$^1$\footnote{e-mail: {\tt frahm@itp.uni-hannover.de}}\\
$^1${\sl Institut f\"ur Theoretische Physik, Universit\"at Hannover \\
D-30167~Hannover, Germany}\\
$^2${\sl Department of Physics, Theoretical Physics, Oxford
University\\ 1 Keble Road, Oxford OX1 3NP, Great Britain}\\
\vfill 
ABSTRACT
\end{center}

\begin{quote}
\baselineskip=14pt 
We study the effects of an integrable impurity in a periodic $t$-$J$
chain. The impurity couples to both spin and charge degrees of freedom
and has the interesting feature that the interaction with the bulk can
be varied continuously without losing integrability. 
We first consider ground state properties close to half-filling in the
presence of a small bulk magnetic field. We calculate the impurity
contributions to the (zero temperature) susceptibilities and the low
temperature specific heat and determine the high-temperature
characteristics of the impurity. We then investigate transport
properties by computing the spin and charge stiffnesses at zero
temperature. Finally the impurity phase--shifts are calculated and the
existence of an impurity bound state in the holon sector is
established. 
\end{quote}

\setlength{\baselineskip}{13pt}

\vfill
PACS-numbers: 71.27.+a~\ 
75.10.Lp~\ 
05.70.Jk~\ 

\vfill
\setcounter{footnote}{0}
\end{titlepage}

\baselineskip=12pt
\section{\sc Introduction}
Theoretical investigations of strongly correlated electron systems have
shown that the low temperature properties of such one dimensional systems
have to be described in terms of a Luttinger liquid rather than a Fermi
liquid. Of particular interest also from an experimental point of view are
the transport properties of these systems in the presence of boundaries and
potential scatterers. 
Several attempts have been made to describe such a situation: the transport
properties of a 1D interacting electron gas in the presence of a potential
barrier have been first studied using renormalization group techniques
\cite{lupe:74,matt:74}.  Triggered by a more recent study of this problem
by Kane and Fisher \cite{kafi:92} different approaches such as boundary
conformal field theory \cite{car:89} and an exact solution by means of a
mapping to the boundary sine-Gordon model \cite{fesa:95,felsa,saleur,tsv0}
have been applied to this problem. In particular the low temperature
properties of magnetic (Kondo) impurities in a Luttinger liquid
\cite{aff:90,afflu:91,frjo:95,frjo:96} have been investigated in great
detail.  In the present work we will investigate the effects of a
particular type of potential impurity in a Luttinger liquid (where both
spin-and charge degrees of freedom are gapless) by means of an exact
solution through the Quantum Inverse Scattering Method (QISM). Attempts to
study effects due to the presence of impurities in many-body quantum system
in the framework of integrable models have a long successful history
\cite{ajo:84,vewo:92,miami,bares:94,kondo,mck}.  As far as lattice models
are concerned the basic mechanisms underlying these exact solutions are
based on the fact that the QISM allows for the introduction of certain
``inhomogeneities''into vertex models without spoiling integrability. Two
approaches are possible and have been studied for various models:\\
Since the ${\cal L}$-operators defining the \emph{local} vertices satisfy a
Yang-Baxter equation with an ${\cal R}$ matrix depending on the difference
of spectral parameters only, one can build families for vertex models with
site-dependent shifts of the spectral parameters. This has been widely used
in solving models for particles with an internal degree of freedom by means
of the nested Bethe Ansatz \cite{yang:67} and, more recently, also for the
construction of systems with integrable impurities \cite{bares:94}
(for a particularly simple case see \cite{schm}). \\
A second possibility is to introduce an impurity by choosing an ${\cal
L}$-operator intertwining between \emph{different} representations of the
underlying algebra (possibly with an additional shift of the spectral
parameter). This has first been used by Andrei and Johannesson to study the
effect of a spin $S$ site in a spin-${1\over2}$ Heisenberg chain
\cite{ajo:84}. Later this approach was used to study chains with
alternating spins \cite{vewo:92,miami}.

Our work is based on the second approach. A novel feature as compared
to \cite{ajo:84} is the presence of a continuously varying free
parameter describing the strength of the coupling of the impurity to
the host chain. The presence of the free parameter is based on
properties of the underlying symmetry algebra of the model, which in
our case is the graded Lie algebra $gl(2|1)$ (see below).

Recently, vertex models and the corresponding quantum chains invariant
under the action of graded Lie algebras have attracted considerable
interest \cite{yue,schlott,eks:92a,ek,eks,martins1,martins2}. In
addition to the well known supersymmetric $t$--$J$ model \footnote{For
a collection of reprints see \cite{rep}.}
\cite{suth:75,schl:87,bares:91,esko:92,foka:93} (which is based on the
fundamental three-dimensional representation of $gl(2|1)$) a model for
electrons with correlated hopping has been constructed using the
one-parametric family of four-dimensional representations of $gl(2|1)$
\cite{bglz:94,bglz:95,befr:95d,bariev,ghlz}. 

In this paper we study the properties of the supersymmetric $t$--$J$ model
with one vertex replaced by an ${\cal L}$ operator acting on a
four--dimensional quantum space. This preserves the supersymmetry of
the model but at the same time lifts the restriction of no double
occupancy present in the $t$--$J$ model at the impurity site. The free
parameter associated with the four--dimensional representation of the
superalgebra allows one to tune the coupling of the impurity to the host
chain. Note that the present models allows for the study a situation 
which in some respects is more general than the ones mentioned above:
the impurity introduced here couples to {\sl both} spin- and charge
degrees of freedom of the bulk Luttinger liquid. The extension of our
calculation to the case of many impurities is straightforward (and
will in fact be used in some parts of the paper).

The paper is organized as follows: In the following section we give a brief
review of the Quantum Inverse Scattering Method and the Bethe Ansatz for
$gl(2|1)$-invariant models. Then we study the ground state properties, low
temperature specific heat and transport properties and show how
they are affected by the presence of the impurity. Finally we compute the
phase shifts acquired by the elementary excitations, holons and
spinons, when scattered off the impurity. 
\section{\sc Construction of the model}
The impurity model is constructed by means of the graded version of
the Quantum Inverse Scattering Method \cite{vladb,Kul,SklKul}.
We start with the ${\cal R}$-matrix of the supersymmetric $t$-$J$
model \cite{esko:92} 
\be
{\cal R}_{tJ}={\cal R}_{33}=\frac{\lambda}{\lambda+i}\Pi +
\frac{i}{\lambda+i} 1\ , 
\label{eq:Rtj}
\ee
with $\Pi^{a_1b_1}_{a_2b_2}=
\delta_{a_1b_2}\delta_{a_2b_1}(-1)^{\epsilon_{b_1}\epsilon_{b_2}}$
being the graded permutation operator acting in the tensor product of
two ``matrix spaces'' isomorphic to the three-dimensional ``quantum
space'' space spanned by $(\uparrow,\downarrow,0)$ with the respective
grading $\e_\uparrow=\e_\downarrow=1$ and $\e_0=0$. 
The corresponding $t$-$J$ ${\cal L}$-operator is given by
${\cal L}_{33}=\Pi {\cal R}_{33}$ and satisfies the following (graded)
intertwining relation
\begin{center}
\setlength{\unitlength}{1cm}
\begin{picture}(7,3)
\put(0,1.5){\line(2,1){2}} \put(0,2){\line(2,-1){2}}
\put(1.8,0.75){\line(-1,5){0.4}} \put(6,2){\line(-2,-1){2}}
\put(6,1.5){\line(-2,1){2}} \put(4.2,0.75){\line(1,5){0.4}}
\put(3.1,1.75){$=$} \put(-2.2,1.70){${\cal R}_{33}(\lambda-\mu)$}
\put(6.3,1.70){${\cal R}_{33}(\lambda-\mu) \quad \ .$}
\put(1.65,2.7){${\cal L}_{33}(\lambda)$} \put(1.75,0.5){${\cal
L}_{33}(\mu)$} \put(4.7,2.7){${\cal L}_{33}(\lambda)$}
\put(4.4,0.5){${\cal L}_{33}(\mu)$}
\end{picture}
\end{center}
\vspace{-0.5cm} 
In components this equations reads
\be
{\cal R}_{33}(\lambda-\mu)_{a_2 c_2}^{a_1 c_1} {\cal L}_{33}
(\lambda)_{\a \gamma}
^{c_1 b_1}{\cal L}_{33}(\mu)_{\gamma \beta}^{c_2 b_2}
(-1)^{\e_{c_2}(\e_{c_1}+\e_{b_1})}=
{\cal L}_{33}(\mu)_{\a \gamma}^{a_1 c_1}
{\cal L}_{33}(\lambda)_{\gamma \beta}^{a_2 c_2}(-1)^
{\e_{a_2}(\e_{a_1}+\e_{c_1})}
{\cal R}_{33}(\lambda-\mu)_{c_2 b_2}^{c_1 b_1}\ .
\label{ir}
\ee

\begin{figure}[ht]
\begin{center}
\leavevmode
\epsfxsize=0.4\textwidth
\epsfbox{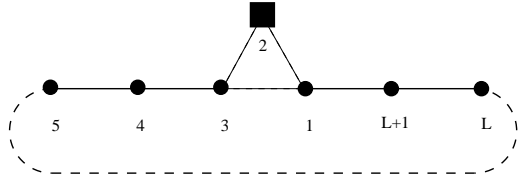}
\end{center}
\vspace{-50mm}
\caption{\label{fig:halg}
The circles denote the normal $t$-$J$-sites. The square denotes the
impurity placed at the second site.}
\end{figure}

The monodromy matrix ${\cal T}_{tJ}$ is defined on the graded tensor product
on quantum spaces and its matrix elements are given by
\be
{\cal T}_{tj}(\lambda)^{a \alpha_1 \cdots \alpha_L}_{b \beta_1 \cdots \beta_L}=
{\cal L}_{33}(\lambda)^{ac_L}_{\alpha_L \beta_L}{\cal L}_{33}
(\lambda)^{c_L c_{L-1}}_{\alpha_{L-1}
\beta_{L-1}} \cdots {\cal L}_{33}(\lambda)^{c_2 b}_{\alpha_1 \beta_1}
(-1)^{\sum_{j=2}^L\left(\epsilon_{\alpha_j}
+\epsilon_{\beta_j}\right)\sum_{i=1}^{j-1}\epsilon_{\alpha_i}}\ .
\ee
The hamiltonian of the $t$-$J$ model is then given as the logarithmic
derivative of the transfer matrix $\tau(\lambda)=\mbox{str}[{\cal
T}_{tJ}(\lambda)]:=\sum_{a=1}^3(-1)^{\epsilon_a} [{\cal
T}_{tJ}(\lambda)]^{aa}$ at zero spectral parameter \cite{esko:92} 
\be {\cal
H}_{tJ}=-i\frac{\partial}{\partial
\l}\ln(\tau_{33})\Big|_{\l=0}-2{\cal N}_e
=-\sum_{k=1}^L(\Pi_{k,k+1}-1) -2{\cal N}_e\ .  \ee 
The ${\cal R}_{33}$-matrix can be constructed as the intertwiner of
the three--dimensional fundamental representation of the superalgebra
$gl(2|1)$. An impurity model can then be constructed in a way
analogous to the spin-$S$ impurity in a spin-$\frac{1}{2}$ Heisenberg
chain \cite{ajo:84} by considering the intertwiner of the
three--dimensional representation with the typical four--dimensional 
representation corresponding to an impurity site with four possible
states ($\up,\da,\up\da=2,0$). In this way we obtain an ${\cal L}$-operator 
${\cal L}_{34}$ with the same auxiliary space dimension as the ${\cal
R}_{33}$-matrix satisfying the following equation:
\begin{center}
\setlength{\unitlength}{1cm}
\begin{picture}(7,3)
\put(0,1.5){\line(2,1){2}} \put(0,2){\line(2,-1){2}}
\put(6,2){\line(-2,-1){2}} \put(6,1.5){\line(-2,1){2}}
\put(1.8,0.75){\line(-1,5){0.4}} \put(1.9,0.75){\line(-1,5){0.4}}
\put(4.1,0.75){\line(1,5){0.4}} \put(4.2,0.75){\line(1,5){0.4}}
\put(3.1,1.75){$=$} \put(-2.2,1.70){${\cal R}_{33}(\lambda-\mu)$}
\put(6.3,1.70){${\cal R}_{33}(\lambda-\mu) \quad \ .$}
\put(1.65,2.7){${\cal L}_{34}(\lambda)$} \put(1.77,0.5){${\cal
L}_{34}(\mu)$} \put(4.7,2.7){${\cal L}_{34}(\lambda)$}
\put(4.4,0.5){${\cal L}_{34}(\mu)$}
\end{picture}
\end{center}
\vspace{-0.5cm} The double line denotes the four--dimensional space
$(\uparrow,\downarrow,2,0)$ with the respective grading
$\e_\uparrow=\e_\downarrow=1$ and $\e_2=\e_0=0$. So double occupancy
is possible at the impurity.  The ${\cal L}_{34}$-matrix is given by
\be
{\cal L}_{34}=\frac{\l-i({\a \over 2}+1)}{\l+i({\a \over 2}+1)}
+\frac{i}{\l+i({\a \over 2}+1)}{\cal L}\ ,
\ee
where ${\cal L}$, expressed in terms of projection operators (the so
called `Hubbard Operators') $X^{ab}=|a\rangle\langle b|$ with
$a,b=\uparrow,\downarrow,2,0$, is given by  {
\begin{equation}
 {\cal L}= 
  \left( \begin{array}{ccc} 
  X_2^{\downarrow \downarrow}+X_2^{00} &
  - X_2^{\downarrow \uparrow} & 
  \sqrt{\a+1}X_2^{0 \uparrow}-\sqrt{\a}X_2^{\downarrow 2} \\
  - X_2^{\uparrow \downarrow} & 
  X_2^{\uparrow \uparrow}+X_2^{00} & 
  \sqrt{\a}X_2^{\uparrow 2}+\sqrt{\a+1}X_2^{0 \downarrow} \\
  \sqrt{\a+1}X_2^{\uparrow 0}-\sqrt{\a}X_2^{2 \downarrow} &
  \sqrt{\a}X_2^{2 \uparrow}+\sqrt{\a+1}X_2^{\downarrow 0} & \a+
  X_2^{\uparrow \uparrow}+X_2^{\downarrow \downarrow}+2 X_2^{00}\\
\end{array} \right) \ ,
\end{equation}}
The parameter $\a>0$ is associated with the four--dimensional
representation of $gl(2|1)$\footnote{Other parameter regions may be
possible, but will not be considered here.} \cite{bglz:94,maas:95}.
By inserting an ${\cal L}_{34}$-operator at one site (for example site
2) we arrive at the following monodromy matrix for the impurity model 
(see Fig.~\ref{fig:halg})
\be
{\cal T}_{imp}(\lambda)^{a \alpha_1 \cdots \alpha_{L+1}}_{b \beta_1 
\cdots \beta_{L+1}}=
{\cal L}_{33}(\lambda)^{ac_{L+1}}_{\alpha_{L+1} \beta_{L+1}}
{\cal L}_{33}(\lambda)^{c_{L+1} c_{L}}_{\alpha_{L} 
\beta_{L}} \cdots 
{\cal L}_{34}(\lambda)^{c_3 c2}_{\alpha_2 \beta_2}
{\cal L}_{33}(\lambda)^{c_2 b}_{\alpha_1 \beta_1}
(-1)^{\sum_{j=2}^{L+1} \left(\epsilon_{\alpha_j}+\epsilon_{\beta_j}\right)
\sum_{i=1}^{j-1}\epsilon_{\alpha_i}}
\ee
The hamiltonian is given by the logarithmic derivative of the
transfer matrix at zero spectral parameter 
\be 
{\cal H}_{tJ-Imp}=-i\frac{\partial \ln(\tau(\lambda))}{\partial
\lambda} |_{\l=0} \ .  
\label{hamil}
\ee 
Due to the fact that ${\cal L}_{34}$ has no shift-point the
hamiltonian contains three-site interactions of the impurity with the
two neighbouring sites. This makes the local hamiltonian rather
complicated. However, as is shown in Appendix C by direct computation,
the hamiltonian constructed in the way described above is invariant
under the graded Lie-algebra $gl(2|1)$. Also in the continuum limit
only a comparably small number of terms relevant in the
renormalization group sense will survive. \hfill\break
The impurity contributions to the hamiltonian are found to be
\be
1-\left({\cal L}_{34}^{-1}(0)\right)^{\alpha_1 \gamma_1}_{\alpha_2 \gamma2}
\Pi^{\gamma_1 \delta_1}_{\alpha_3  \beta_3}
\left({\cal L}_{34}(0)\right)^{\delta_1 \beta_1}_{\gamma_2 \beta_2}
(-1)^{[(\epsilon_{\alpha_3}+\epsilon_{\beta_3})\epsilon_{\gamma 2}]}
-i\left({\cal L}_{34}^{-1}(0)\right)^{\alpha_1 \gamma_1}_{\alpha_2 \gamma2}
\left({\cal L}_{34}'(0)\right)^{\gamma_1 \beta_1}_{\gamma_2 \beta_2}\ ,
\ee
which can be expressed in terms of the Hubbard operators as
\begin{eqnarray}
{\cal H}_{1,2,3} &=& 1+\frac{\a+1}{({\a \over 2}+1)^2}-{1 \over 
({\alpha \over 2}+1)^2}  \times \nonumber \\
&&\Big[ 
 \tilde{\Pi}_{12} +\a X_1^{00}+H_{12}+{\a^2 \over 4}\left(1-2 X_1^{00}\right)
\Pi_{13}(1-2 X_1^{00}) \nonumber \\ 
&&+\left(\tilde{\Pi}_{23}+X_2^{22}(1+\Pi_{13})\right)
 -{\a \over 2}\left[\left(1-2 X_1^{00}\right)\Pi_{13}\tilde{\Pi}_{12} +h.c.
 \right] \nonumber \\
&&-{\a \over 2}\left[\left(1-2 X_1^{00}\right)\Pi_{13}H_{12} +h.c.\right]
 + H_{12} \Pi_{13} H_{12}+\left(H_{12}\Pi_{13}\tilde{\Pi}_{12}+h.c.\right)
\Big] \ .\nonumber 
\end{eqnarray}
Here $\tilde{\Pi}_{12}$ denotes a modified permutation operator
\be 
\tilde{\Pi}_{12}=\left(-X_2^{22}+(1-X_2^{22})\Pi_{12}(1-X_2^{22})\right).
\ee
The operator $H_{12}$ is given by
{\small
\bea
H_{12}&=& 
X_1^{\up 0}\left[(\sqrt{\a+1}-1) X_2^{0 \uparrow}
-\sqrt{\a}X_2^{\downarrow 2}\right]
+X_1^{\da 0}\left[(\sqrt{\a+1}-1) X_2^{0 \downarrow}
+\sqrt{\a}X_2^{\uparrow 2}\right]+{ h.c.}
\eea
}
{From} the explicit form of the hamiltonian it is clear that the
impurity can be thought of as being attached from the outside to the
$t$-$J$ chain as is depicted in Fig.~\ref{fig:halg}

\begin{figure}[ht]
\begin{center}
\leavevmode
\epsfxsize=0.4\textwidth
\epsfbox{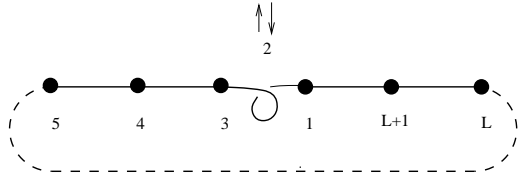}
\end{center}
\vspace{-5cm}
\caption{\label{fig:hainf}
Hamiltonian in the limit $\a \to \infty$. As in the half filled case the
double occupied impurity site decouples of the bulk-chain which has to
be solved with twisted boundary conditions.}
\end{figure}

\begin{figure}[ht]
\begin{center}
\leavevmode
\epsfxsize=0.4\textwidth
\epsfbox{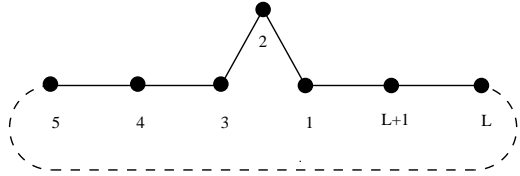}
\end{center}
\vspace{-5cm}
\caption{\label{fig:hnull}
Effective ground state hamiltonian in the limit $\a \to 0$. The
impurity site is just an extra site of the $t$-$J$ model. (This is
correct for all particle densities below half filling where the
impurity site becomes doubly occupied.)}
\end{figure}

In the limiting cases $\a \to \infty$ and $\a \to 0$ the impurity
contribution simplifies essentially
\bea 
\lim_{\a \to 0} H_{1,2,3} &=&
(1-n_{2,\uparrow}n_{2,\downarrow})(2-\Pi_{12}-\Pi_{23})
(1-n_{2,\uparrow}n_{2,\downarrow})+
n_{2,\uparrow}n_{2,\downarrow}(3-\Pi_{13}) , \nn 
\lim_{\a \to \infty} H_{1,2,3} &=&
1-(1-2 X_1^{00})\Pi_{13}(1-2X_1^{00}).
\label{hgl}
\eea
It is shown below that for $\a=0$ the impurity site behaves like an
``ordinary'' $t$-$J$ site in the ground state below half-filling. In
the limit $\a=\infty$ the impurity becomes doubly occupied and causes
the hopping amplitude between the two neighbouring $t$-$J$ sites to
switch sign. The situation is equivalent to a decoupled impurity and a
$t$-$J$ chain with $L$ sites and a twist of $-1$ in the boundary
conditions. These two situations are shown in Fig.~\ref{fig:hainf} and
Fig.~\ref{fig:hnull} respectively.

\begin{figure}[ht]
\begin{center}
\leavevmode
\epsfxsize=0.4\textwidth
\epsfbox{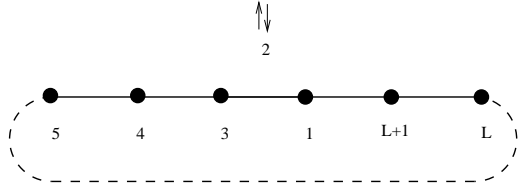}
\end{center}
\vspace{-5mm}
\caption{\label{fig:hhalb}
Decoupling of the double occupied impurity-site of the bulk-chain
at half--filling.}
\end{figure}

For the half--filled case (two electrons at the impurity site) the
impurity decouples from the host chain and we
obtain the following hamiltonian (see Fig.~\ref{fig:hhalb})
\be
H_{1,2,3}^{{1 \over 2} - filling} = -\Pi_{13}+1+\frac{4}{\a+2} ,
\label{hgr}
\ee 

If we wish to consider the case of many impurities the above
construction will go through as long as every ${\cal L}$-operator 
${\cal L}_{34}$
is sandwiched by two ${\cal L}$-operators ${\cal L}_{33}$. The resulting 
hamiltonian will consist of a sum over terms of the form $H_{k,k+1,k+2}$ 
and $1-\Pi_{jj+1}$. The maximal number of impurities is thus bounded by
the number of $t$-$J$ sites.
\section{\sc Algebraic Bethe Ansatz}
The Bethe Ansatz equations can be derived as in \cite{esko:92}.
The monodromy matrix is a $3 \times 3$ matrix of quantum operators
acting on the entire chain
\be
{\cal T}_{imp}=
\left( 
  \begin{array}{ccc}
   A_{11}(\l) & A_{12}(\l) & B_{1}(\l)  \\
   A_{21}(\l) & A_{22}(\l) & B_{2}(\l)  \\
   C_{1}(\l) & C_{2}(\l) & D(\l)
  \end{array}
 \right)\ .
\ee
Like in the case of the $t$-$J$ model one constructs the eigenstates
of the hamiltonian by successive application of the $C_i$-operators on
the bosonic reference state $|0\rangle$ 
\be
|\l_1,\dots,\l_n|F\rangle=C_{a_1}(\l_1)C_{a_2}(\l_2)
\dots C_{a_n}(\l_n)|0\rangle F^{a_n \dots a_1}
\label{state}
\ee
The values of the spectral parameters $\l_i$ and the coefficients
$F^{a_n \dots a_1}$, which are constructed by means of a nested
algebraic Bethe ansatz (NABA), are determined by requiring the
cancellation of the so-called ``unwanted terms''. 
For the construction of the NABA one only needs the intertwining relation
(\ref{ir}) (and the existence of the reference state).
Due to the fact that the corresponding ${\cal R}$-matrix is the same
for the impurity model and the $t$-$J$ model the resulting equations
are very similar. The only modification of the Bethe equations arises
from a different eigenvalue of the $A_{ii}$-operators on the reference
state leading to the following form for the Bethe equations
\begin{eqnarray}
 \left( \frac{\l_j-{i\over 2}}{\l_j+{i\over 2}} \right)^{L} \left(
 \frac{\l_j-{\alpha+1 \over 2}i}{\l_j+{\alpha+1 \over 2}i} \right) &=&
 \prod^{N_\downarrow}_{\alpha=1} \frac
 {\l_j-\lambda^{(1)}_\alpha-{i\over 2}}
 {\l_j-\lambda^{(1)}_\alpha+{i\over 2}}\ , \quad
 j=1,\ldots,N_e=N_\uparrow+N_\downarrow \ , \nonumber \\
 \prod^{N_e}_{j=1} \frac{\lambda^{(1)}_\alpha-\l_j+{i\over 2}}
 {\lambda^{(1)}_\alpha-\l_j -{i\over 2}} &=& -
 \prod^{N_\downarrow}_{\beta=1}
 \frac{\lambda^{(1)}_\alpha-\lambda^{(1)}_\beta+i}
 {\lambda^{(1)}_\alpha-\lambda^{(1)}_\beta-i}\ , \quad
 \alpha=1,\ldots,N_\downarrow \ .
\label{eq:bagi}
\end{eqnarray}
The energy of a Bethe state of the form \r{state} with spectral
parameters $\{\l_j,\ j=1\dots N_e\}$, $\{\l^{(1)}_\g\ , \g=1\dots
N_\da\}$ in the grand canonical ensemble is given by
\be
E=\sum_{j=1}^{N_e}{1 \over \l_j^2+{1\over 4}} 
-\mu N_e -H \frac{N_\uparrow-N_\downarrow}{2}\ .  
\ee
Here $\mu$ is the chemical potential and $H$ is a bulk magnetic field.
Properties of the ground state and excitations as well as the
thermodymanics can now as usual be determined {\sl via} an analysis of
the Bethe equations \r{eq:bagi}.
In order to analyze ground state and excitations we first have to
discuss some details concerning the lattice length. We note that the
length of the lattice is $L+1$, where $L$ is the length of the $t$-$J$
``host'' chain. It is known that the ground state of the $t$-$J$ model
changes if the length of the lattice is increased by one. The
situation is similar in the present model. A unique antiferromagnetic
ground state exists for odd (equal) numbers of up and down spins. If
we require a smooth limit to the half-filled band with a doubly
occupied impurity we find that the length of the host chain must be of
the form $L=0\ {\rm mod}\ 4$. 

If we consider the case of $n_i$ impurities and $N$ $t$-$J$ sites
discussed above the left-hand-side of the first set of Bethe equations
changes to $\left( \frac{\l_j-{i\over 2}}{\l_j+{i\over 2}} \right)^{N}
\left(  \frac{\l_j-{\alpha+1 \over 2}i}{\l_j+{\alpha+1 \over 2}i}
\right)^{n_i}$ whereas the other equations remain unchanged.
\section{\sc Ground State Properties}
In order to study ground state properties we need to know the
configuration of $\l_j$'s and $\l^{(1)}$'s corresponding to the lowest
energy state for given $\mu$ and $H$. It can be found in complete
analogy to the $t$-$J$ model without impurity:
for finite magnetic field the ground state is described in terms of
two filled Fermi seas of
\begin{itemize}
\item{} complex $\l$-$\l^{(1)}$- ``strings'' $\l_\pm=\l^{(1)} \pm
\frac{i}{2}$ \cite{schl:87} (where $\l^{(1)}$ are real and fill all
vacancies between the Fermi rapidity $B$ and $\infty$ and $-\infty$
and $-B$). As we approach half-filling $B$ tends to $0$. Removing one
$\l$-$\l^{(1)}$-``string'' from the Fermi sea and placing it outside
leads to a particle-hole excitation involving only charge degrees of
freedom (``holon-antiholon'' excitation)\cite{bares:91}. 

\item{} real solutions $\l_j$ associated with the spin degrees of
freedom. They are filling all vacancies between
$A$ and $\infty$ and $-\infty$ and $-A$. We note that
$A\rightarrow\infty$ as $H\rightarrow 0$.
\end{itemize} 
At zero temperature the dressed energies $\Psi(\l)$ and
$\varepsilon(\l)$ of the excitations associated with charge and spin
degrees of freedom respectively have to satisfy the following coupled
integral equations \cite{schl:87}
\bea 
\Psi(\l) &=& 2\pi a_2(\l)-2\mu-\int_{B}^{\infty} \!\!\!\!\!\!+\!\!\!
\int_{-\infty}^{-B} d\mu\ a_2(\l-\mu)
\Psi(\mu)-\int_{A}^{\infty}\!\!\!\!\!\!+\!\!\! \int_{-\infty}^{-A}
d\mu\ a_1(\l-\mu) \varepsilon(\mu) , \nn 
\varepsilon(\l) &=& 2\pi
a_1(\l)-\mu-\frac{H}{2}-\int_{B}^{\infty}\!\!\!\!\!\!+\!\!\!
\int_{-\infty}^{-B} d\mu\ a_1(\l-\mu) \Psi(\mu) 
\eea
where $a_n(\l) = \frac{1}{2\pi} \frac{n}{\l^2+\frac{n^2}{4}}$.  
The functions $\Psi$ and $\varepsilon$ are negative in the intervals
$(-\infty,B) \cup (B,\infty)$ and $(-\infty,A) \cup (A,\infty)$ as
they are monotonically decreasing functions of $|\l |$.

The ground state energy per site is given by 
\be 
\frac{E-\mu N_e -HS^z}{L}=
-\Psi(0)-2\mu+2+\frac{1}{L}\left(-2\mu+2-\frac{2 \alpha}{\alpha+2} -
\int_{-B}^B d\mu\ a_\a(\mu) \Psi(\mu) \right).
\label{energy}
\ee 
The first part on the r.h.s. is the contribution of the $L$ sites of
the host $t$-$J$ chain, whereas the second term is the
contribution from the impurity. Transforming the integral equations
into the complementary integration intervals we obtain the following
equations
\bea 
\eps_s(\l)&=& -2\pi
a_1(\l)+H-\int_{-A}^{A}d\mu\ a_2(\l-\mu)\ \eps_s(\mu) + \int_{-B}^B
d\mu\ a_1(\l-\mu)\ \eps_c(\mu)\ ,\nn 
\eps_c(\l)&=& \mu-\frac{H}{2}+\int_{-A}^{A}d\mu\ a_1(\l-\mu)\
\eps_s(\mu)\ ,
\label{dden2}
\eea 
where $\eps_s=-\eps$ and $\eps_c=-\Psi$. The ground state energy per
site is now given by
\bea 
\frac{E-\mu N_e -HS^z}{L}&=& \eps_c(0)-2\mu+2+\frac{1}{L} \left(
\eps_{\alpha} +\frac{4}{\alpha+2} - 2\mu \right)\ ,
\label{Egs}
\eea
where 
\bea 
\eps_{\alpha}=\int_{-B}^B d\l\ \eps_c(\l) a_{\alpha}(\l)\ .
\eea
We now split \r{Egs} into two parts: the contribution of the $t$-$J$
host chain $e_{host}$ and the contribution of the impurity $E_{imp}$.
Note that in \r{Egs} we divided by the length of the host chain
instead of the length of the lattice. The impurity contribution to the
ground state energy is then given by
\be
E_{\rm imp} = \left(\eps_{\alpha} +\frac{4}{\alpha+2} - 2\mu
\right). 
\label{Eimp}
\ee

Before we turn to an analysis of the impurity contributions to the
ground state energy we give a brief review of the properties of the
$t$-$J$ host chain. The ground state properties of the $t$-$J$ model
have been studied in great detail in \cite{kaya:91,bares:92}: below a
critical density $n_c$, which is related to the magnetic field by 
\be
H=4\sin^2\left({\pi n_c \over 2}\right),
\label{hc}
\ee
the integration boundary $B$ in \r{dden2} is $\infty$ and the system
behaves like a Fermi-liquid with all spins up. The particle density is
a function of $A$ in this case.
For densities larger than $n_c$ the system is a Luttinger liquid
exhibiting spin and charge separation.
For general band-fillings the integral equations \r{dden2} can be
solved only numerically (in the almost empty band it is possible to
reduce them to a system of coupled Wiener-Hopf equations but we do not
pursue this avenue here). 
For densities slightly above $n_c$ ($B \gg 1$) and $B \gg A$ the integral
equations can be solved analytically. 
Evaluating equation (\ref{dden2}) in this case leads to:
\bea 
\eps_s(\l)&=& -2\pi
a_1(\l)+H+\left(\mu - {H \over 2}\right) \left(1- {1 \over \pi B}\right)
+{\cal O} \left({1 \over B^3}\right)\ ,\nn 
\eps_c(\l)&=& \mu-\frac{H}{2}+\int_{-A}^{A}d\mu\ a_1(\l-\mu)\
\eps_s(\mu)\ ,
\label{Bgga}
\eea  
Via $\eps_s(A)=0$ and $\eps_c(B)=0$ the integration boundaries depend on
the magnetic field and the chemical potential in the following way
\be
A_c=\frac{\sqrt{4-H}}{2 \sqrt{H}}+\cdots , \qquad B_c=\frac{\sqrt{4
\arctan(2A_c) -\sqrt{H}\sqrt{4-H}}}{\sqrt{2 \pi}\sqrt{\mu-H/2}}+\cdots
\ee
Using the densities $\rho_c$ and $\rho_s$ (see (\ref{rn})) 
the zero--temperature density and magnetization per site are given by:
\be
D_{bulk}=1-\left(1-\frac{1}{\pi B_c}\right) \frac{2}{\pi}
\arctan(2A_c) , \qquad 
m_{bulk}=\frac{1}{2}-\left(1+\frac{1}{\pi B_c}\right) \frac{1}{\pi}
\arctan(2A_c)  
\label{tjc}
\ee
Taking the limit $H \to 0$ with finite $B_c$ is not permitted in
(\ref{tjc}) as this would imply $A \to \infty$ in contradiction to $B
\gg A$\footnote{The correct result would be $D_{bulk}={2 \sqrt{\mu}
\over \pi}$ and not $D_{bulk}={\sqrt{\mu} \over \pi}$ (see
\cite{juku:96}).}.

For the almost half-filled band and a small magnetic field \r{dden2}
can be solved analytically as well \cite{schl:87}:
combining a Wiener--Hopf analysis for small magnetic field $(A\gg 1)$
and an iterative solution near half-filling $(B\ll 1)$ one obtains
\cite{es:96}
\bea 
\eps_c(0)= -\mb -2a-2\ln(2) \frac{\zeta(3)}{\pi}B^3 
\eea 
where
$\mb=2\ln(2)-\mu,\ |\mb|\ll 1$ and
\be 
a= {\pi \over e}e^{-2\pi A}=\frac{H^2}{8\pi^2}
(1-\frac{1}{2\ln(H)}+\ldots )\ ,\quad 
B^2=\frac{2}{3\zeta(3)}\left[\mb+2a+\frac{8\ln(2)}{3\pi}
\frac{1}{\sqrt{6\zeta(3)}}(\mb+2a)^{\frac{3}{2}}\right] \ .
\label {eq:a}
\ee
{From} these results we can now determine bulk and impurity
contributions to the ground state energy.
In order to give a reference frame for the impurity properties we
start by giving the results for the (bulk) zero-temperature
magnetization per site, magnetic susceptibility, density and
compressibility:
\bea
m_{bulk}&=& \frac{H}{2\pi^2}\left(1-\frac{1}{2\ln(H)}\right)
\left(1+\frac{\ln(2)}{\pi}\sqrt{\frac{8(\mb+2a)}{3\zeta(3)}}\right)
+\ldots\ ,\nn \chi_{H,bulk}&=&
\frac{1}{2\pi^2}\left(1-\frac{1}{2\ln(H)}\right)\left(1+\frac{\ln(2)}{\pi}
\sqrt{\frac{8(\mb+2a)}{3\zeta(3)}}\right)
\nn &+&
\frac{2\ln(2)}{\pi}\frac{H^2}{4\pi^4}\left(1-\frac{1}{\ln(H)}\right)
\frac{1}{\sqrt{6\zeta(3)(\mb+2a)}}+\ldots\ ,\nn D_{bulk}&=& 1 -
\frac{2\ln(2)}{\pi}\sqrt{\frac{2(\mb+2a)}{3\zeta(3)}}+\ldots\ ,\nn
\chi_{c,bulk}&=&
\frac{2\ln(2)}{\pi}\frac{1}{\sqrt{6\zeta(3)[\mb+2a]}}+\ldots\ .
\label{bulk}
\eea

The physical implications of \r{bulk} have been discussed extensively
in the literature \cite{kaya:91,bares:92}. We merely note the
divergences in the susceptibilities as we approach half-filling.
\subsection{\sc Ground State Properties of the Impurity}
The impurity contribution to the ground state energy is of the same
form as the surface energy of an open $t$-$J$ chain with boundary
chemical potential studied in \cite{es:96} (more precisely the
impurity contribution is of the same form as the boundary chemical
potential dependent part of the surface energy).
The parameter $\a$ plays a role similar to the boundary chemical
potential in the open chain. The leading order impurity contributions
for small bulk magnetic fields close to half-filling can thus be
calculated in the same way as in \cite{es:96} with the result
\bea 
\eps_{\alpha} &=&
\cases{-\frac{4\zeta(3)}{\pi\a}B^3+\ldots
& {\rm if} $ \alpha \gg B$ \cr \eps_c (0)\left(1-\frac{2\a}{\pi
B}\right)+\ldots & {\rm if} $\alpha \ll B$ \cr }
\label{epsi}
\eea
Together with \r{Eimp} this gives the leading impurity contribution to
the ground state energy. Note that by differentiating the impurity
contribution to the ground state energy w.r.t. $\a$ it is possible to
evaluate the expectation values of various operators at the
impurity. However due to the complicated structure of these operators
it is difficult to extract useful information from the expectation
values and we therefore do not present these calculations here.

For densities below and slightly above $n_c$ analytical results for
particle number and magnetization can be obtained as well, whereas for
generic values of $\a$ the integral equations can be solved only
numerically. 
Below we first present the aforementioned analytical results and then
turn to the numerical solution of the integral equations.
\subsubsection{\sc Large $\a$ close to half-filling}
By taking the appropriate derivatives of the impurity contribution to
the ground state energy we can evaluate the impurity contribution to
the magnetization, particle number and susceptibilities. On physical
grounds it is reasonable to assume that the impurity contributions to
magnetization and particle number are concentrated in the vicinity of
the impurity (see also Appendix B). We find 
\bea 
M_{\alpha}&=&\frac{H}{\pi^3\a}\left(1-\frac{1}{2\ln(H)}\right)
\sqrt{\frac{8(\mb+2a)}{3\zeta(3)}}+\ldots\ ,\nn 
\chi_{H,\alpha}&=&\frac{1}{\pi^3\a}\sqrt{\frac{8(\mb+2a)}{3\zeta(3)}}
+\frac{H^2}{\pi^5\a}(1-\frac{1}{\ln(H)})\frac{1}{\sqrt{6\zeta(3)(\mb+2a)}}
+\ldots\ ,\nn 
N_{\alpha}&=&2-\frac{4}{\pi\a}\sqrt{\frac{2(\mb+2a)}{3\zeta(3)}}+\ldots\
,\nn 
\frac{\partial N_{\alpha}}{\partial
\mu}&=&\frac{4}{\pi\a}\frac{1}{\sqrt{6\zeta(3)[\mb+2a]}}+\ldots 
\label{agr}
\eea 
By inspection of \r{agr} we find that close to half-filling the
impurity is on average almost doubly occupied and thus only weakly
magnetized. The susceptibilities exhibit the same types of
singularities as the bulk.

In comparison to the bulk the following ratios are found: 
\bea
\frac{M_{\alpha}}{m_{bulk}}&=& \frac{4}{\pi\alpha}
\sqrt{\frac{2(\mb+2a)}{3\zeta(3)}}\ ,\quad
\frac{\chi_{H,\alpha}}{\chi_{H,bulk}}= \cases{
\frac{4}{\pi \alpha}\sqrt{\frac{2(\mb+2a)}{3\zeta(3)}} & if $H^2\ll\mb$\cr
\frac{2}{\alpha \ln(2)} & $ \longrightarrow \frac{1}{2}$-filling }\ .  
\eea

\begin{figure}[ht]
\begin{center}
\leavevmode
\epsfxsize=0.4\textwidth
\epsfbox{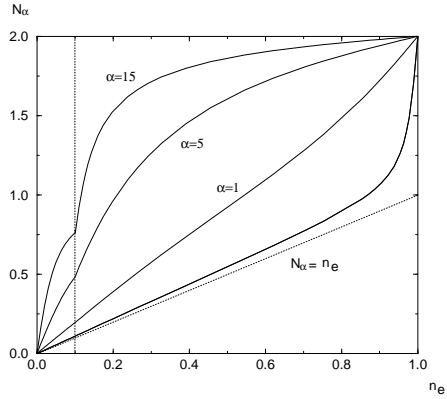}
\end{center}
\vspace{-1cm}
\caption{\label{fig:ni}
Number of electrons located at the impurity as a function of
the bulk electron--density for fixed magnetic field $H=0.1$ and 
several values of $\a$. The dotted line denotes the critical electron
density  $(4.7)$ corresponding to $H$.}
\end{figure}

The analytic (\ref{agr}) and numerical (Fig.~\ref{fig:ni}) results
show that the impurity is doubly occupied in the limit $\a \to
\infty$. Thus the effective ground state hamiltonian in this case is a
$t$-$J$ model of $L$ sites with twisted boundary conditions (see
(\ref{hgl})) as represented by Fig.~\ref{fig:hainf}. This is
consistent with the twisted boundary conditions occuring in
(\ref{eq:bagi}) in this limit.
\subsubsection{\sc Small $\a$ close to half-filling}
For small values of $\a$ such that $\a\ll B$ we find the following
results for the impurity contributions to magnetization, particle
number and susceptibilities 
\bea M_{\alpha}&=&
\frac{H}{2\pi^2}\left(1-\frac{1}{2\ln(H)}\right)
\left(1-\frac{\alpha}{\pi B}\right)
\left(1+\frac{2 \ln(2)}{\pi}\sqrt{\frac{2(\mb+2a)}{3\zeta(3)}}\right)
+\ldots\ ,\nn 
\chi_{H,\alpha}&=&
\frac{1}{2\pi^2}\left(1-\frac{1}{2\ln(H)}\right)(1-\frac{\alpha}{\pi B})
+ \frac{H^2}{4\pi^4}\left(1-\frac{1}{\ln(H)}\right)\frac{\alpha}{2 \pi
B}\frac{1}{(\mb+2a)}+\ldots, \nn 
N_{\alpha}&=& 1+\frac{\alpha}{\pi B}-
(1-\frac{\alpha}{\pi B})\frac{2\ln(2)}{\pi}
\sqrt{\frac{2(\mb+2a)}{3\zeta(3)}}+\ldots\ ,\nn 
\frac{\partial N_{\alpha}}{\partial \mu} &=& (1-\frac{\alpha}{\pi
B})\frac{2\ln(2)}{\pi}
\frac{1}{\sqrt{6\zeta(3)[\mb+2a]}}+\frac{\alpha}{2 \pi
B}\frac{1}{(\mb+2a)}+\ldots 
\label{akb}
\eea 
Comparing this to (\ref{bulk}) we see that the impurity behaves
almost like an ``ordinary'' site of the lattice and the model reduces
to a $t$-$J$ chain on $L+1$ sites in the limit $\a \to 0$ (see (\ref{hgl})) 
as represented by Fig.~\ref{fig:hnull}. 
\subsubsection{\sc Densities below the critical electron density $n_c$}
Below the critical electron--density (\ref{hc}) the particle number at
the impurity--site is given by (see Fig.~ \ref{fig:ni})
\be
N_\a(n_e)=1-\frac{2}{\pi}\arctan\left(\frac{\cot({\pi n_e \over 2})}{1+\a}
\right)
\label{idifer}
\ee
In this regime only the states $|0\rangle $ and $| \uparrow \rangle$ 
are realized in the ground state, hence the hamiltonian ${\cal H}_{1,2,3}$ 
simplifies to
\bea
{\cal H}_{1,2,3}&=&X_3^{\uparrow \uparrow}+X_1^{\uparrow \uparrow}+
{2 \over 1+{\a \over 2}}X_2^{\uparrow \uparrow} \nn
&-&{1 \over 1+{\a \over 2}} \left[-{\a \over 2}
\left(X_1^{\uparrow 0}X_3^{0 \uparrow}+h.c.\right)
+\sqrt{\a+1}\left(X_1^{\uparrow 0}X_2^{0 \uparrow}+
+X_2^{\uparrow 0}X_3^{0 \uparrow}+h.c.\right)\right]
\eea
\subsubsection{\sc Densities slightly above  the critical electron density 
$n_c$}
Slightly above $n_c$ the impurity contributions to magnetization
and particle number are found to be
\bea
N_\a &=& 1-\frac{2}{\pi}\arctan\left(\frac{2 A_c}{1+\a}\right)
\left(1-\frac{1}{\pi B_c}\right)+\frac{\a}{\pi B_c}+\ldots ,\nn
M_\a &=& \frac{1}{2}-\frac{1}{\pi}\arctan\left(\frac{2 A_c}{1+\a}\right)
\left(1+\frac{1}{\pi B_c}\right)-\frac{\a}{2 \pi B_c}+\ldots 
\label{ncc}
\eea
Combining (\ref{ncc}) and (\ref{tjc}) we obtain the following result
\be
{\partial N_\a \over \partial n_e}|_{n_e=n_c^+}=\frac{\a+1-N_\a(n_c)}
{1-n_c}\ .
\label{abnc}
\ee
\subsubsection{\sc Numerical Results for general band filling and
magnetic field}
For general magnetic fields and band fillings the impurity
magnetization and particle number can be determined by numerically
solving the relevant integral equations. 

\begin{figure}[ht]
\begin{center}
\leavevmode
\epsfxsize=0.4\textwidth
\epsfbox{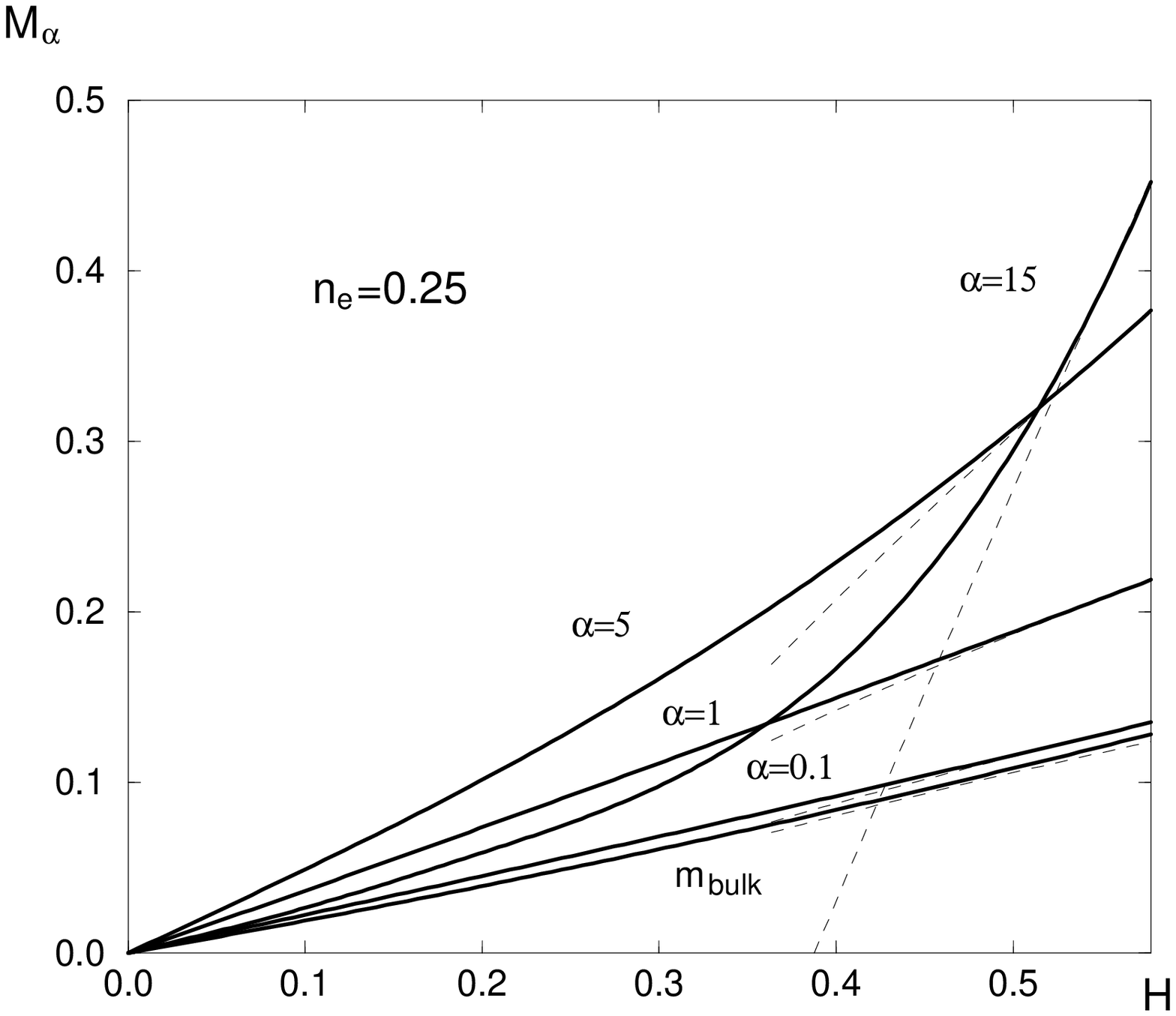}
\end{center}
\vspace{-1cm}
\caption{\label{fig:impmag1}
Impurity magnetization as a function of magnetic field for band
filling $0.25$ and  several values of $\a$. The dashed lines denote
the asymptotic behaviour (4.20)  
near the critical magnetic field.}
\end{figure}

\begin{figure}[ht]
\begin{center}
\leavevmode
\epsfxsize=0.4\textwidth
\epsfbox{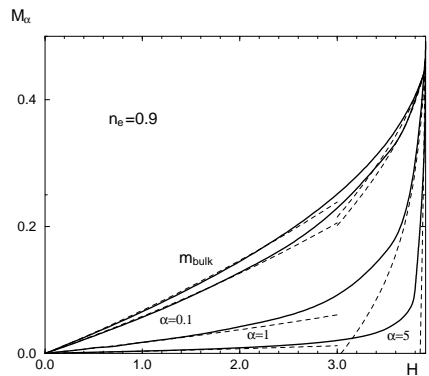}
\end{center}
\vspace{-1cm}
\caption{\label{fig:impmag2}
Impurity magnetization as a function of magnetic field for band
filling $0.9$ and  several values of $\a$. The dashed lines denote
the asymptotic behaviour for small magnetic field 
(see (4.15) and (4.17)) and near the critical magnetic field (4.20).
}
\end{figure}

Some results for the impurity magnetization as a function of the
magnetic field for different values of $\a$ and band fillings are
shown in Fig. \ref{fig:impmag1} and Fig. \ref{fig:impmag2}. 

Only magnetic fields below the critical values are considered. For
comparison the analytical results obtained above are included in the
figures.
The shape of the magnetization curves is entirely different from the
ones obtained in the Kondo model where (as a function of $H$) there is
a crossover from a linear behaviour for small fields to a constant
value at large fields.

\begin{figure}[ht]
\begin{center}
\leavevmode
\epsfxsize=0.4\textwidth
\epsfbox{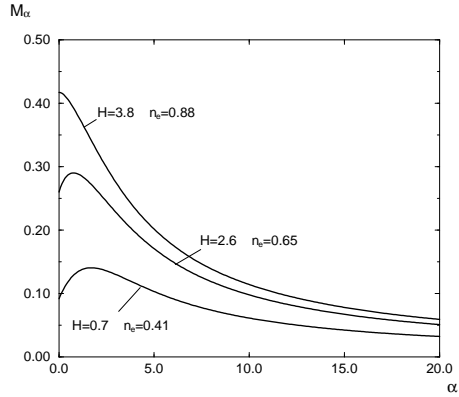}
\end{center}
\vspace{-1cm}
\caption{\label{fig:nhfest}
Impurity magnetization as a function of $\a$ for several constant
magnetic fields and electron densities.}
\end{figure}

{}From (\ref{idifer}) one finds that the impurity magnetization at the
critical magnetic field $H_c$ is an increasing function of $\a$.
Numerical results show that this behaviour persists for $H<H_c$, small
$\a$ and not too large densities (see Fig.~\ref{fig:impmag1} and
Fig.~\ref{fig:nhfest}). On the other hand the impurity is almost
doubly occupied and hence only weakly magnetized for large $\a$ and
magnetic fields sufficiently smaller than $H_c$. This leads to a
decrease of the magnetization with $\a$, which for small densities
and small magnetic field fields is shown in Fig.~\ref{fig:impmag1} and
Fig.~\ref{fig:nhfest}. Near half--filling the magnetization is a
decreasing function of $\a$ (see (\ref{agr}) and (\ref{akb})) for all 
magnetic fields not too close to $H_c$ (see Fig.~\ref{fig:impmag2} 
and Fig.~\ref{fig:nhfest}).

\begin{figure}[ht]
\begin{center}
\leavevmode
\epsfxsize=0.4\textwidth
\epsfbox{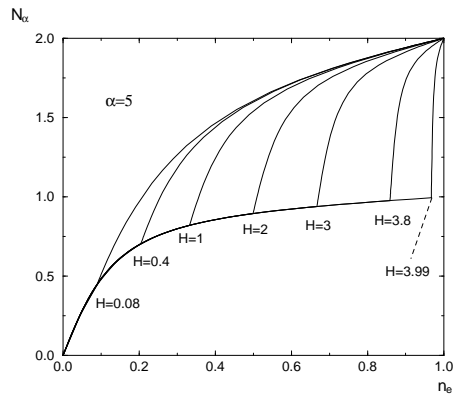}
\end{center}
\vspace{-1cm}
\caption{\label{fig:na5}
Number of electrons located at the impurity as a function of
the bulk electron--density for various values of the magnetic field and 
$\a=5$.}
\end{figure}

The impurity particle number $N_\a$ as a function of band filling for
various values of $\a$ is shown in Fig.~\ref{fig:ni}. 
For small $\a$ $N_\a$ at first increases linearly with band-filling
and then curves upwards, reaching $2$ at half-filling. For very small
$\a$ the deviation from the $N_\a=n_e$-behaviour occurs very close to
half-filling. For large values of $\a$ and densities above $n_c$
$N_\a$ increases rapidly with band-filling, then evens out and
eventually reaches $N_\a=2$. The crossover between these two regimes
occurs roughly for $\a\approx 1$ for small magnetic field .
The magnetic-field dependence of the impurity particle number is shown
in Fig.~\ref{fig:na5}, where $N_\a$ is shown for the case $\a=5$ and
several values of the magnetic field. We see that apart from the shift
in critical density $n_c$ the behaviour of $N_\a$ is qualitatively
unchanged as $H$ is increased. The derivative of $N_\a$ at $n_c$ is
given by (\ref{abnc}).
\subsection{\sc Fermi velocities}
As we will now show the impurity contribution to the susceptibilities
are related to a modification of the Fermi velocities by the
impurity. The bulk Fermi velocities are defined by
\be
v_c=\frac{\eps'_c(B)}{2\pi \rho_c(B)} \ ,\qquad
v_s=\frac{\eps'_s(A)}{2\pi \rho_s(A)}\ ,
\ee
where the root densities $\rho_s(\l)$ and $\rho_c(\l)$ are solutions
of the integral equations
\bea
\rho_s(\l)&=&  a_1(\l)-\int_{-A}^{A}d\mu\ a_2(\l-\mu)\ \rho_s(\mu)
+ \int_{-B}^B d\mu\ a_1(\l-\mu)\ \rho_c(\mu)\ ,\nn
\rho_c(\l)&=& \int_{-A}^{A}d\mu\ a_1(\l-\mu)\ \rho_s(\mu)\ .
\label{rn}
\eea
The Fermi velocities can be calculated for small magnetic field near
half filling (using the same techniques as above) leading to the
results
\be
v_c=\frac{3\zeta(3)}{2\ln(2)}B+... ,\qquad 
v_s=\pi\left(1-\frac{2\ln(2)B}{\pi}-\frac{1}{(2\ln(H/H_0))^2}\right)+... 
\ee
where $H_0=\sqrt{8\pi^3/e}$.
Near the critical electron density $n_c$ we obtain
\be
v_c=\frac{2}{B_c}-\frac{4 A_c}{(1+4A_c^2)\arctan(2A_c)B_c}+... ,\qquad 
v_s=\frac{8 A_c}{1+4A_c^2}+... 
\ee
The respective Fermi velocities of the entire system (bulk+impurity)
$\tilde{v}_c$ and $\tilde{v}_s$ are given by
\be
\tilde{v}_\b=\frac{\eps'_\b(A_\b)}{2\pi \left(\rho_\b(A_\b)+{1 \over
L}\rho^{(1)}_\b(A_\b)\right)}=\frac{v_\b}{1+{1 \over
L}{\rho^{(1)}_\b(A_\b)\over \rho_\b(A_\b)}}\ ,\quad \b=s,c\ ,
\label{fv}
\ee
where $A_s=A$, $A_c=B$ and $\rho^{(1)}_\b$ satisfy the following
integral equations
\bea 
\rho_s^{(1)}(\l)&=&
-\int_{-A}^{A}d\mu\ a_2(\l-\mu)\
\rho_s^{(1)}(\mu) +\int_{-B}^{B}d\mu\ a_1(\l-\mu)\
\rho_c^{(1)}(\mu) \nn \rho_c^{(1)}(\mu)&=& a_\a (\mu)
+\int_{-A}^{A}d\mu'\ a_1(\mu-\mu')\ \rho_s^{(1)}(\mu')\
\eea
For later convenience we define the ratios
\be
f_\b=\frac{{\rho}^{(1)}_\b(A_\b)}{\rho_\b(A_\b)} \ .
\label{fb}
\ee
In the bulk the Fermi velocities are related to the susceptibilities
{\sl via} \cite{bares:92}
\bea
\chi_{c,bulk}&=&\frac{1}{\pi
n_e^2}\left[\frac{(Z_{cc})^2}{v_c}+\frac{(Z_{cs})^2}{v_s}\right] , \nn
\chi_{H,bulk}&=&\frac{1}{4\pi}\left[\frac{(Z_{cc}-2Z_{sc})^2}{v_c}+
\frac{(Z_{cs}-2Z_{ss})^2}{v_s}\right].
\label{chiba}
\eea
Here $Z_{\a \b}$ are the elements of the so-called 
dressed charge matrix (see e.g. \cite{woyn:89})
\begin{equation}
{\bf Z}= \left( \begin{array}{cc} Z_{cc} & Z_{sc} \\ 
  Z_{cs} & Z_{ss} \end{array} \right)=
  \left( \begin{array}{cc} \xi_{cc}(B) & \xi_{sc} (B) \\ 
  \xi_{cs} (A) & \xi_{ss} (A) \end{array} \right) \ ,
\label{eq:z}
\end{equation}
where $\xi_{\a\b}(\l)$ are solutions of the integral equations
\begin{equation}
\left( \begin{array}{cc} \xi_{cc}(\vartheta) & \xi_{sc}(\vartheta) \\ 
  \xi_{cs}(\lambda) & \xi_{ss}(\lambda) \end{array} \right)=
\left( \begin{array}{cc} 1 & 0 \\ 0 & 1 \end{array} \right)+
 \left( \begin{array}{cc} 0 & \int_{-A}^{A}d\mu\ a_1(\vartheta-\mu) \\
\int_{-B}^{B}d\mu\ a_1(\l-\mu) &  
-\int_{-A}^{A}d\mu\ a_2(\l-\mu) \end{array} \right)*
 \left( \begin{array}{cc} \xi_{cc}(\mu) & \xi_{sc}(\mu) \\ 
  \xi_{cs}(\mu) & \xi_{ss}(\mu) \end{array} \right)\ .
\end{equation}
Replacing the velocities in (\ref{chiba}) by $\tilde{v}_c$ and
$\tilde{v}_s$ we see that the impurity contribution to the magnetic
susceptibility is given by 
\bea
\chi_{H,\a}&=&\frac{1}{4\pi L}\left[\frac{(Z_{cc}-2Z_{sc})^2f_c}{v_c}
+\frac{(Z_{cs}-2Z_{ss})^2f_s}{v_s}\right].
\label{chihi}
\eea
There are two impurity contributions to the compressibility: one is due to
the change of the electron density $n_e$, the other to the modification of
the Fermi velocities
\be
\chi_{c,\a}=-\frac{2N_\a}{D_{bulk} L} \chi_{c,bulk}+\frac{1}{\pi
n_e^2 L}\left[\frac{(Z_{cc})^2f_c}{v_c}+\frac{(Z_{cs})^2f_s}{v_s}\right].
\label{chici}
\ee
The second part can be identified with ${1 \over L n_e^2}
\frac{\partial N_{\alpha}}{\partial \mu}$.

Let us consider the two cases for which we presented analytical
results above in more detail:
\begin{itemize}
\item{$A \gg 1, B \ll 1$} corresponding to a small bulk magnetic field
and an almost half-filled band. The leading contributions to the
dressed charge matrix are given by 
{\small
\bea
Z_{cc} = 1+\frac{2\ln(2)}{\pi}B + \ldots \quad &,& \quad  
Z_{sc}=\left[{1 \over
2}-\frac{\sqrt{2a}}{\pi}\left(1-\frac{1}{4\ln(H)-2\ln(8\pi^3/e)}\right)
\right]\left[1+\frac{2\ln(2)}{\pi}B\right]+ \ldots , \nn
Z_{cs} = 2B\sqrt{a} \quad &,& \quad
Z_{ss}=\frac{1}{\sqrt{2}}\left(1-\frac{1}{4\ln(H/H_0)}\right)
+B\sqrt{a}+\ldots
\label{dcl}
\eea
}
where $A,B,a$ are given by \r{eq:a}. With aid of the Wiener-Hopf
technique we can obtain the ratios $f_c$ and $f_s$ as well.  For $\a
\gg B$ they are given by
\be
f_c=\frac{2}{\ln(2) \a}+\frac{2a}{\ln^2(2)\a} +\ldots , \quad
f_s=\frac{4B}{\pi \a} +\ldots\ .
\ee
Inserting these results in (\ref{chihi}) and (\ref{chici}) we reproduce 
the results (\ref{agr}) (to ${\cal O}({1 \over \ln(H)})$).
For $\a \ll B$ we find
\be
f_c=1+\frac{\a}{2 \ln(2) B^2}\left(1+{a \over \ln(2)}\right)+\ldots , 
\quad f_s=1-\frac{\a}{\pi B}+\ldots
\ee
which leads to the same resluts as (\ref{akb}). 

\item{$0<n_e-n_c\ll 1$} corresponding to densities slightly above the
critical density $n_c$. We find
\bea
Z_{cc} = 1+\left(1-{1 \over \pi B_c}\right) {A_c \over \pi B_c^2}+ \ldots 
\quad &,& \quad  
Z_{sc}= {A_c \over \pi B_c^2}+ \ldots , \nn
Z_{cs} = 1- {1 \over \pi B_c}\quad &,& \quad
Z_{ss}=1+\ldots. 
\label{dcnc}
\eea
The ratios $f_\b$ are given by
\be
f_c=\frac{2 \arctan \left({2 A_c \over \a+1}\right)+\pi \a}{2 \arctan(2A_c)},
\quad f_s=\frac{a_{1+\a}(A_c)}{a_1(A_c)}. 
\ee
\end{itemize}
\section{\sc The Impurity at Finite Temperatures}
The thermodynamics of the supersymmetric $t$-$J$ model was studied by
Schlottmann in \cite{schl:87}. The TBA equations for the dressed
energies are the same in the presence of the impurity in complete
analogy with {\sl e.g.} \cite{ajo:84}, so that we can simply quote the
result from \cite{schl:87}
\bea
\epsilon&=&2\pi a_1-\mu-\frac{H}{2}+Ta_1*\ln(1+e^{-\frac{\Psi}{T}})
-T \sum_{n=1}^\infty a_n*\ln(1+e^{-\frac{\phi_n}{T}})\nn
\Psi&=&-2\mu +2\pi a_2+Ta_1*\ln(1+e^{-\frac{\epsilon}{T}})
+T a_2*\ln(1+e^{-\frac{\Psi}{T}})\nn
\phi_n&=&nH-T\ln(1+e^{-\frac{\phi_n}{T}})+
Ta_n*\ln(1+e^{-\frac{\epsilon}{T}})
+T \sum_{m=1}^\infty \theta_{nm}*\ln(1+e^{-\frac{\phi_m}{T}})\ ,
\label{tba}
\eea
where
\be
\theta_{nm}(\l)=\frac{1}{2\pi}\int_{-\infty}^\infty d\omega
e^{-i\omega\lambda}\coth(|\frac{\omega}{2}|)
\left(e^{-|\frac{\omega}{2}(m-n)|}-e^{-|\frac{\omega}{2}(m+n)|}\right).
\ee
The bulk free energy is given by \cite{schl:87}
\be
F_{\rm bulk}= -\Psi(0)-2\mu ,
\ee
whereas the impurity contribution to the free energy can be cast in
the form
\be
F_{\rm imp}=-2\mu-\frac{2\a}{\a+2}-\int_{-\infty}^{\infty}d\l\ 
a_\a(\l)\left[\Psi(\l)+T\ln(1+e^{-\frac{\Psi(\l)}{T}}\right].
\ee
We note that in the zero temperature limit this reproduces correctly
the ground state energy  \r{energy}.
In the high-temperature limit the TBA equations \r{tba} turn into
algebraic equations that can be solved by Takahashi's method
\cite{taka:71}. The leading terms of the high-temperature expansion are
given by
\bea
\frac{F_{\rm bulk}}{L}&=&-T
\ln\left(1+e^{\frac{\mu}{T}}2\cosh\frac{H}{2T}\right) \nn
F_{\rm imp}&=& -\frac{2\a}{\a+2}-T
\ln\left(1+e^{\frac{2\mu}{T}}+e^{\frac{\mu}{T}}2\cosh\frac{H}{2T}\right).
\eea
We see that these yield the correct values of the entropy of a system
of $L$ sites with $3$ degrees of freedom and one site with four
degrees of freedom in the limit $T\rightarrow\infty$. We also note
that the parameter $\a$ enters only in a trivial way into the leading
term of the high-temperature expansion. By taking the appropriate
derivatives we can compute the mean values of particle number and
magnetization 
\bea
\langle D_{\rm bulk}\rangle&=&\frac{2\cosh\frac{H}{2T}}
{\exp(-\frac{\mu}{T})+2\cosh\frac{H}{2T}}\ ,\quad 
\langle m_{\rm bulk}\rangle=\frac{\sinh\frac{H}{2T}}
{\exp(-\frac{\mu}{T})+2\cosh\frac{H}{2T}}\ ,\nn
\langle N_{\rm imp}\rangle&=&
\frac{\exp(\frac{\mu}{T})+\cosh\frac{H}{2T}}
{\cosh(\frac{\mu}{T})+\cosh\frac{H}{2T}}\ ,\quad 
\langle M_{\rm imp}\rangle =
\frac{\sinh\frac{H}{2T}}
{\cosh(\frac{\mu}{T})+\cosh\frac{H}{2T}}\ .
\eea
Half-filling corresponds to the limit $\mu\rightarrow\infty$, in which
the impurity is on average doubly occupied and unmagnetized whereas
the bulk exhibits a magnetization per site of
$\frac{1}{2}\tanh\frac{H}{2T}$. 
\section{\sc Low Temperature Specific Heat}
\begin{figure}[ht]
\begin{center}
\leavevmode
\epsfxsize=0.4\textwidth
\epsfbox{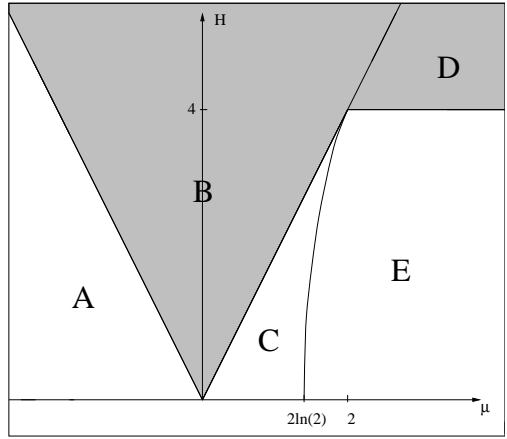}
\end{center}
\vspace{-5mm}
\caption{
\label{fig:phase}
The phase diagramm of the $t$-$J$ model, shaded regions correspond to 
a ferromagnetic ground state.}
\end{figure}

In order to calculate the contributions of the impurity to the
low--temperature specific heat we need to consider different
characteristic regions (see Fig. \ref{fig:phase}) of the $t$-$J$
model as it was done for the Hubbard model by Takahashi \cite{tak:74}.
As in \cite{tak:74} we assume $H \gg T$ so we can neglect 
the effects of the string solutions. (An alternative approach which
overcomes this restriction has recently been used in \cite{juku:96} to
compute bulk thermodynamic properties in the $t$-$J$ model).
\\
{\bf Region A:} The region is characterized by $\mu < -{H \over 2}$. 
For $T=0$ the electron density is zero as $A=B=\infty$. The low temperature
free energy per site is given by 
\bea
\frac{\Delta F_{\rm bulk}}{L}= \frac{F_{\rm bulk}-F_{\rm
bulk}(T=0)}{L}&=&-T \int_{-\infty}^{\infty}d\l\ a_2(\l)\ln(1+e^{-\Psi
\over T}) -T \int_{-\infty}^{\infty}d\l\ a_1(\l)\ln(1+e^{-\varepsilon
\over T}) \nn 
&\approx& -{2 T \over \pi} \int_{0}^{\infty}d\l\ {1 \over \l^2}
\ln(1+e^{2 \mu \over T}e^{-2 \over  \l^2 T})
-{T \over \pi} \int_{0}^{\infty} d\l\ {1 \over \l^2}\ln(1+e^{\mu + H/2\over T}
e^{-1 \over  \l^2 T}) \nn
&\approx& -{1 \over \sqrt{2 \pi}}T^{3/2} e^{2 \mu \over T}
-{1 \over 2 \sqrt{\pi}}T^{3/2} e^{\mu +H/2 \over T}
\eea 
The impurity contribution to the low temperature free energy is
\be
\Delta F_\a=F_{\rm imp}-F_{\rm imp}(T=0)=-T \int_{-\infty}^{\infty}
d\l\ a_\a(\l)\ln(1+e^{-\Psi \over T}) \approx  -{\a \over 2 \sqrt{2
\pi} }T^{3/2} e^{2 \mu \over T}\ .
\ee
\\
{\bf Region B:} For $T=0$ this region, characterized by $-H/2 < \mu < H/2$, 
is the ferromagnetic phase with electron density varrying as $0 < n_e < 1$.
The right boundary line is defined by $B=\infty$ the left one by $A=B=\infty$.
The low temperature free energies are given by 
\be
\frac{\Delta F_{\rm bulk}}{L} \approx -{\pi T^2 \over 6}{1 \over v_s} \qquad
\Delta F_\a \approx -{\pi T^2 \over 6}{f_s \over v_s}\ .
\ee
\\
{\bf Region C:} In this region, characterized by $0< A,B < \infty $, the 
electron density varries between $0 < n_e < 1$ at zero temperature. 
The right boundary line defined by $B=0$ can be calculated with aid 
of (\ref{eq:a}) for small magnetic field and with an iterative solution 
of (\ref{dden2}) for $H \stackrel{<}{\sim} 4$.
The low temperature free energies are given by 
\be
\frac{\Delta F_{\rm bulk}}{L}\approx -{\pi T^2 \over 6}
\left[ {1 \over v_s}+{1 \over v_c} \right] \qquad
\Delta F_\a \approx -{\pi T^2 \over 6}
\left[ {f_s \over v_s}+{f_c \over v_c} \right]
\ee
\\
{\bf Region D:} This region is characterized by $\mu > H/2 > 2$.
For $T=0$ we obtain the ferromagnetic half--filled band. The low temperature
free energy is given by 
\be
\frac{\Delta F_{bulk}}{L} = -T \int_{-\infty}^{\infty}d\l\
a_1(\l)\ln(1+e^{-\varepsilon_s \over T}) \approx -{4 T \over \pi}
\int_{0}^{\infty}d\l\ \ln(1+e^{4-H \over T}e^{-16 \l^2 \over T})
\approx -{1 \over 2 \sqrt{\pi}}T^{3/2} e^{4-H \over T}\ .
\ee
The impurity contribution to the low temperature free energy is
\be
\Delta F_\a=-T \int_{-\infty}^{\infty}d\l\ a_\a(\l)
\ln(1+e^{-\varepsilon_c \over T})\approx  -T e^{H/2-\mu \over T}\ .
\ee
\\
{\bf Region E:} For $T=0$ this region is half-filled and non ferromagnetic.
The low temperature free energy of the bulk is given by 
\be
\frac{\Delta F_{bulk}}{L} \approx-{\pi T^2 \over 6}{1 \over v_s}
\ee
The impurity contribution in this region can not be calculated in closed form.
In the two limiting cases $A \ll 1$ and $A \gg 1$ we obtain the
following results
\be
\Delta F_\a\approx 
-{2 \over \a}T^{3/2}
\left\{
     \begin{array}{ll}
       {1 \over \sqrt{3\pi \zeta(3)/2-\pi^3a}} e^{2\ln(2)-\mu+2a \over T} &
        \hbox{for~} A \gg 1 \\[8pt]
       \sqrt{3 \over 8}{1 \over (4-H)^{{3 \over 4}}} e^{H/2-\mu+{2 \over 3 \pi}
        (4-H)^{3/2} \over T} &
				 \hbox{for~} A \ll 1
     \end{array}\right.
\ee
\\
{\bf Wilson ratio in Region C:} The specific heat of the bulk and the
impurity in region C are given by 
\be
C_{v,bulk}=\frac{\pi T}{3}\left(\frac{1}{v_s}+\frac{1}{v_c}\right)
\qquad
C_{v,\a}=\frac{\pi T}{3L}\left(\frac{f_s}{v_s}+\frac{f_c}{v_c}\right)
\label{cv}
\ee
In deriving these results we assumed that $H\gg T$. Defining
\be
R=\frac{\chi_{H,\a} / \chi_{H,bulk}}{C_{v,\a}/C_{v,bulk}}
=\frac{\left[\left(\frac{Z_{cc}-2Z_{sc}}{Z_{cs}-2Z_{ss}}\right)^2
\frac{f_c}{f_s}+\frac{v_c}{v_s}\right]\left[1+\frac{v_c}{v_s}\right]}
{\left[\left(\frac{Z_{cc}-2Z_{sc}}{Z_{cs}-2Z_{ss}}\right)^2
+\frac{v_c}{v_s}\right]\left[\frac{f_c}{f_s}+\frac{v_c}{v_s}\right]}.
\ee
and if we assume that the limits $T\rightarrow 0$ and $H\rightarrow 0$ 
commute \footnote{This holds in all known cases for the specific heat in
integrable spin chains where the ground state contains only real roots
of the Bethe equations. The assumption is also supported by the
findings of \cite{juku:96} where it was shown to be true for the bulk
specific heat.} we can calculate a ``Wilson ratio''
\be
R_W=\lim_{H \to 0} R=\frac{1+\frac{v_c}{v_s}}
{\frac{f_c}{f_s}+\frac{v_c}{v_s}}\ .
\ee
Unlike for the case of the Kondo model (where spin and charge degrees
of freedom decouple and the impurity couples only to the spin) there is
no reason to expect $R_W$ to be universal for the present model.

\begin{figure}[ht]
\begin{center}
\leavevmode
\epsfxsize=0.4\textwidth
\epsfbox{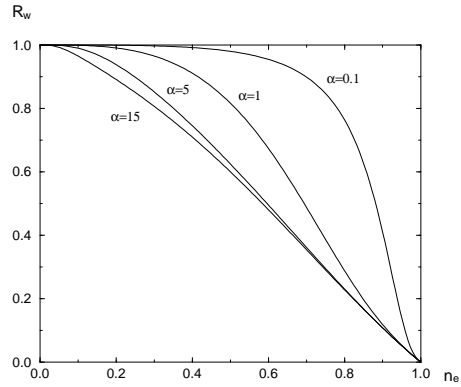}
\end{center}
\vspace{-1cm}
\caption{
\label{fig:rw}
Wilson--ratio as a function of the electron--density.}
\end{figure}

{}From our analytical calculations we find for $B\ll 1$, $B \ll \a$ that
$R_W=1-n_e\ll 1$. 
Near the empty band the ratio $R_W$ tends to one because $v_c$ approaches
zero. Numerical calculations show that the limiting value one is most
rapidly approached for small values of $\a$ (see Fig.~\ref{fig:rw}).
For comparison we quote the result for a Kondo impurity in a Luttinger
liquid \cite{frjo:96}, for which $R_W=\frac{4}{3}(1+\frac{v_s}{v_c})$.
\section{\sc Transport Equations}
Following Shastry and Sutherland \cite{shas:90} (see also \cite{zyv})
we will now calculate spin- and charge stiffnesses from the
finite-size corrections to the ground state energy of the model with
twisted boundary conditions. For the $t$-$J$ model the bulk
stiffnesses were determined in \cite{kaya:91}. Our analysis follows
closely the discussion given in \cite{bares:92}.
Our goal is to evaluate the ground state energy as a function of the
twist angles $\phi_\uparrow$ and $\phi_\downarrow$. Imposing twisted
boundary conditions the BAE (\ref{eq:bagi}) are modified in the
following way
\begin{eqnarray}
 \left( \frac{\l_j-{i\over 2}}{\l_j+{i\over 2}} \right)^{L} 
 \left( \frac{\l_j-{\alpha+1 \over 2}i}{\l_j+{\alpha+1 \over 2}i} \right) &=&
 e^{-i \phi_\uparrow} \prod^{N_\downarrow}_{\gamma=1} 
 \frac {\l_j-\lambda^{(1)}_\gamma-{i\over 2}}
 {\l_j-\lambda^{(1)}_\gamma+{i\over 2}}\ ,
 \quad j=1,\ldots,N_e \ ,
 \nonumber \\
 \prod^{N_e}_{j=1} \frac{\lambda^{(1)}_\gamma-\l_j+{i\over 2}}
 {\lambda^{(1)}_\gamma-\l_j -{i\over 2}} &=&
  - e^{i(\phi_\uparrow-\phi_\downarrow)}\prod^{N_\downarrow}_{\beta=1}
 \frac{\lambda^{(1)}_\gamma-\lambda^{(1)}_\beta+i}
 {\lambda^{(1)}_\gamma-\lambda^{(1)}_\beta-i}\ ,
 \quad \gamma=1,\ldots,N_\downarrow  \ .
\label{eq:bagitwi}
\end{eqnarray}
For technical reasons it is convenient to use a different representation
of the BAE first introduced by Sutherland for the $t$-$J$ model
without impurity \cite{suth:75}. This can be obtained by a
particle--hole transformation in the space of rapidities which is done
in Appendix A. The final BAE are given by
\footnote{\noindent
Considering the half--filled case $N_h=0$ we see that the impurity
leaves the BAE (\ref{eq:bagist}) unchanged, which is not
immediately obvious from the other set of BAE (\ref{eq:bagitwi}).}
{\small
\begin{eqnarray}
 \left( \frac{\l_j+{i\over 2}}{\l_j-{i\over 2}} \right)^{L} 
  &=& -e^{i\phi_s}
 \prod^{M^{(1)}+N_\downarrow-1}_{k=1} \frac {\l_j-\l_k+i} {\l_j-\l_k-i}\ 
 \prod^{M^{(1)}}_{\g=1}
 \frac{\l_j-\l_\g^{(1)}-{i\over 2}}{\l_j-\l_\g^{(1)}+{i\over 2}},
 \quad j=1,\ldots,  L+1-N_\uparrow\ ,
 \nonumber \\
 \prod^{M^{(1)}+N_\downarrow-1}_{j=1}
 \frac{\l_\g^{(1)}-\l_j-{i\over 2}}{\l_\g^{(1)}-\l_j+{i\over 2}}
 &=& e^{-i\phi_c} 
 \frac{\l_\g^{(1)}+i{\alpha \over 2}}{\l_\g^{(1)}-i{\alpha \over 2}}
 \quad \g=1,\ldots,M^{(1)}=L+2-N_\uparrow-N_\downarrow  \ ,
\label{eq:bagist}
\end{eqnarray}
}
where the twist--angles are given by
$\phi_s=\phi_\uparrow-\phi_\downarrow$ and $\phi_c=\phi_\downarrow$
\cite{bares:92}.

The equations \r{eq:bagist} can be simplified by making use of the
`string-hypothesis'\footnote{Note that we do not consider string
solutions to the BAE in order to determine the stiffnesses and the
results are thus independent of the precise form of the strings.},
which states that for $L\rightarrow\infty$ all solutions are composed
of real $\1l_\g$'s whereas the $\l$'s are distributed in the complex
plane according to the description
\be
\l^{n,j}_\b = \l^{n}_\b + i\left(\frac{n+1}{2}-j\right)\ ,\quad
j=1\ldots n\ , 
\label{strings}
\ee
where $\b=1\ldots M_n$ labels different `strings' of length $n$ and
$\l^{n}_\b$ is real. The imaginary parts of the $\l$'s can now be
eliminated from \r{eq:bagist} via \r{strings}. Taking the logarithm of
the resulting equations (for $M_n$ strings \r{strings} of length $n$
and $M^{(1)}$ $\l^{(1)}$'s (note that $\sum_{n=1}^\infty nM_n=
L+1-N_\uparrow$)) we arrive at 
\bea
\frac{2\pi}{L} I^n_\b &=& \theta(\frac{\l^n_\b}{n}) -
\frac{1}{L} \sum_{(m\gamma)}\theta_{mn}(\l^n_\b
-\l^m_\g) + \frac{1}{L}
\sum_{\gamma=1}^{M^{(1)}}\theta(\frac{\l^n_\b-\1l_\g}{n})
+\frac{\phi_s}{L}, \b=1\ldots M_n\nn
\frac{2\pi}{L} J_\g &=& \frac{1}{L}\sum_{(n\b)}\theta(\1l_\g-\l^n_\b)
+ \frac{1}{L}\theta(\frac{\1l_\g}{\a})+\frac{\phi_c}{L}\ , \g=1\ldots
M^{(1)}\ ,
\label{baelog}
\eea
where $I^n_\b$ and $J_\g$ are integer or half-odd integer numbers,
$\theta(x)=2\arctan(2x)$ and
\be
\theta_{n,m}(x) = (1-\delta_{m,n})\theta ({x\over{|n-m|}}) +
2\ \theta ({x\over{|n-m|+2}})+\dots +2\ \theta ({x\over{n+m-2}}) +
\theta ({x\over{n+m}})\ .
\ee
For vanishing twist angles the ranges of the ``quantum numbers''
$I^{n}_\b$ and $J_\g$ are given by
\bea
|I^n_\b|&\leq& \frac{1}{2}(L+M^{(1)}+M_n-2\sum_{m=1}^\infty {\rm
min}\{m,n\}M_m-1) \ ,\nn
|J_\g|&\leq& \frac{1}{2}(\sum_{n=1}^\infty M_n -1)\ .
\label{int}
\eea
Ground state and excitations can now be constructed by specifying sets
of integer (half-odd integer) numbers $I^n_\b$ and $J_\g$ and turning
equations \r{baelog} into sets of coupled integral
equations in the thermodynamic limit. The antiferromagnetic ground
state for zero twist angles is obtained by filling consecutive quantum
numbers $I^1_\b$ and $J_\g$ symmetrically around zero, which
corresponds to filling two Fermi seas of spin and charge degrees of
freedom respectively. 
The effect of an infinitesimally small flux is to shift the
distribution of roots ({\sl i.e.} the rapidities in the Fermi seas) by
a constant amount. This shift leads to a twist-angle dependent
contribution to the ground state energy. 
The ground state (in the presence of flux) is described in terms of
the root densities $\rho_\a(\l)$, which are solutions of the integral
equations 
\bea
\rho_s(\l)&=& a_1(\l)-\int_{\tilde{\La}_s^-}^{\tilde{\La}_s^+}d\mu\
a_2(\l-\mu)\ \rho_s(\mu)
+\int_{\tilde{\La}_c^-}^{\tilde{\La}_c^+}d\mu\ a_1(\l-\mu)\
\rho_c(\mu)+{\cal O}(L^{-2}) \nn
\rho_c(\mu)&=& \int_{\tilde{\La}_s^-}^{\tilde{\La}_s^+}d\mu'\
a_1(\mu-\mu')\ \rho_s(\mu')\ + \frac{a_\a (\mu)}{L}+{\cal O}(L^{-2}).
\label{gss}
\eea
Here $\tilde{\La}_\b^\pm$ are the Fermi points for the finite system
in the presence of the flux. 
We denote the Fermi points for the infinite system {\sl without}
flux by $\pm\La_{\b,0}$ (note that the distribution of
roots is symmetric around zero in that case). For further convenience
we define the quantities 
\be
D_\b = -\frac{1}{2} \int_{\tilde{\La}_\b^+}^{\infty}
\!\!\!\!\!\!-\!\!\! \int_{-\infty}^{\tilde{\La}_\b^-} d\mu\
\rho_\b(\mu) \ .
\ee
In order to evaluate the stiffnesses we need to consider infinitesimal
flux only, which yields a correction of order $\frac{1}{L}$ to the
ground state energy. Following through the standard steps
\cite{woyn:89,frko:90} and taking into account the
$\frac{1}{L}$-corrections from the infinitesimal flux we
obtain\footnote{Similar expressions are obtained for the corrections
to excited state energies.}
\be
E(\phi_s,\phi_c,\alpha)=L
e(\La_{\b,0})+f(\La_{\b,0})-\frac{\pi}{6L}(v_s+v_c)+2\pi L
D^T {\bf Z} {\bf V} {\bf Z}^T D 
+o(\frac{1}{L}),
\label{fscorr}
\ee
where ${\bf Z}$ is the dressed charge matrix (\ref{eq:z}),
$D=(D_c,D_s)^\top$
and $V={\rm diag}(v_c,v_s)$. Here $e(\La_{\b,0})$ and $f(\La_{\b,0})$ are
the ground state energy per site and impurity energy in the infinite
system without flux.
The quantity $D_\b$ is given by
\be
D_\b=\frac{\phi_\b}{2\pi L}\ .
\label{dk}
\ee

In order to study the charge and spin conductivities and the respective
currents we consider the energy-difference $\Delta
E=E(\phi_s,\phi_c,\a)-E(0,0,\a)$. It can be cast in the form
\be
\Delta E=
{1 \over L} \phi_\a D_{\a \b} \phi_\b\ 
+\ {\rm subleading\ terms}.
\label{De}
\ee
According to \cite{shas:90} the charge (spin) stiffness $D^{(\rho)}$
($D^{(\sigma)}$) is defined as
\be
D^{(\rho)}={L \over 2} {\partial^2 \over \partial \phi^2}\Delta
E(\phi_c=\phi,\phi_s=0,\alpha)\ ,\quad
D^{(\sigma)}={L \over 2} {\partial^2 \over \partial \phi^2}\Delta
E(\phi_c=\phi,\phi_s=-2\phi,\alpha).
\label{dd}
\ee
Using the expressions \r{dk} in \r{fscorr} and \r{dd} we find that the
stiffnesses are not modified by the impurity to leading order in
$\frac{1}{L}$. It is clear from the above analysis that there are corrections
due to the impurity in the subleading terms. The precise form of these
expressions is not of particular interest from a physical point of
view and as the extension of the above finite-size analysis to the
subleading orders is rather difficult we refrain from determining
them.
The important point is that despite the presence of the impurity the
stiffnesses are still finite and the dc-conductivity is thus
infinite. We believe that this fact is due to the integrability and
the related absence of backscattering.

This means that the integrable impurity considered here is of a completely
different nature than the ``weak link''-type potential impurity considered
in \cite{lupe:74,matt:74,kafi:92,tsv0}: as the electron-electron
interactions are repulsive in the supersymmetric $t$-$J$ model a weak link
drives the system to a strong coupling fixed point characterized by the
vanishing of the conductivity.
\subsection{\sc Stiffnesses for finite density of impurities}
Let us now turn to transport properties for the system with a finite
density of impurities. The necessary steps are the same as above, the
main difference being the change of integral equations describing the
ground state from \r{gss} to 
\bea
\rho_s(\l)&=&
(1-n_i)a_1(\l)-\int_{\tilde{\La}_s^-}^{\tilde{\La}_s^+}d\mu\ 
a_2(\l-\mu)\ \rho_s(\mu)+\int_{\tilde{\La}_c^-}^{\tilde{\La}_c^+}
d\mu\ a_1(\l-\mu)\ \rho_c(\mu)\ ,\nn
\rho_c(\mu)&=& n_ia_\a (\mu)+\int_{\tilde{\La}_s^-}^{\tilde{\La}_s^+}d\mu'\
a_1(\mu-\mu')\ \rho_s(\mu')\ ,
\label{gss2}
\eea
where $n_i$ is the concentration of impurities.
As the integral equations for the dressed energies remain unchanged,
the dressed charge is not modified and $A(\mu,H)$ and $B(\mu,H)$
are not changed. However, the presence of a finite density of
impurities leads to changes in the electron density, which is now
given by $n_e=(1-n_i)D_{bulk}+n_i N_\a$ and the Fermi velocities,
which are found to be of the form
\be
\tilde{v}_\b=\frac{v_\b}{1+n_i(f_\b-1)}\ .
\ee
Here $v_\b$ are the velocities of the normal $t$-$J$-chain and the 
$f_\b$ are defined in (\ref{fb}). 
{From} (\ref{dd}) the following form of the stiffnesses is easily
deduced
\be
D^{(\rho)}= {1 \over 2 \pi}\left(\tilde{v}_c 
Z^2_{cc}+\tilde{v}_s Z^2_{cs}\right), \qquad
D^{(\sigma)}= {1 \over 2
\pi}\left(\tilde{v}_c (Z_{cc}-2Z_{sc})^2 +\tilde{v}_s
(Z_{cs}-2Z_{ss})^2 \right).
\label{stiff}
\ee
Using the results of the above sections we can evaluate these expressions
analytically for small magnetic field close to ``maximal filling''
($t$-$J$ sites singly occupied, impurities doubly occupied) and near
the critical electron density $n_c$. This is done in the following two
subsections. Finally we present numerical results for the general case.

\subsubsection{\sc Stiffnesses for small magnetic field near maximal
filling}

Close to maximal filling and for $\a \gg B$ the leading term of the
charge stiffness is found to be
\be
D^{(\rho)}\approx {1 \over 2 \pi}\tilde{v}_c Z^2_{cc}=
\frac{3 \zeta(3)}{4 \pi \ln(2)} \frac{B}{1+n_i\left(\frac{2}{\ln(2) \a}
+\frac{2a}{\ln^2(2)\a}-1\right)}+\ldots
\label{drho}
\ee
Combining \r{drho} with the result for the electron density we obtain
the following limiting value for the slope of the charge stiffness as
a function of density
\be
{\partial D^{(\rho)} \over \partial n_e}|_{n_e=1+n_i}=-{3 \zeta(3)
\over 8 \ln^2(2)} {1 \over \left[1+n_i\left(\frac{2}{\ln(2)
\a}+\frac{2a}{\ln^2(2)\a} -1\right)\right]^2}+\ldots
\ee
In the strong-coupling limit this turns into
\be
\lim_{\a \to \infty}{\partial D^{(\rho)} \over \partial
n_e}|_{n_e=1+n_i}= -{3 \zeta(3) \over 8 \ln^2(2)} {1 \over
\left(1-n_i\right)^2} ={1 \over \left(1-n_i\right)^2} {\partial
D^{(\rho)}_{tJ} \over \partial n_e}|_{n_e=1}
\label{dgr}
\ee
The leading term of the spin stiffness for $B \ll 1$ and $\a \gg B$ is given
by 
\be
D^{(\sigma)}\approx{1 \over 2 \pi}\tilde{v}_s
\left(Z_{cs}-2Z_{ss}\right)^2 ={v_s \over \pi}\left(1-{1 \over 4
\ln(H/H_0)}\right)^2  {1 \over 1+n_i(f_s-1)}\ .
\ee
In the limit $\a \to \infty$ we find
\be
{\partial D^{(\sigma)} \over \partial n_e}|_{n_e=1+n_i}=\pi
\left(1-{1 \over 2 \ln(H/H_0)}\right) {1 \over (1-n_i)^2} 
={1 \over \left(1-n_i\right)^2}{\partial D^{(\sigma)}_{tJ} \over
\partial n_e}|_{n_e=1}\ .
\label{dgs}
\ee
\subsubsection{\sc Stiffnesses slightly above $n_c$}
As pointed out above it is possible to derive analytic expressions for
the stiffnesses for densities slightly above $n_c$. However the
resulting expressions are found to be rather complicated so that we
refrain from listing them here.

The derivative of the spin--stiffness with respect to the electron density
at $n_e=n_c^+$ is always positive, taking its maximum at $\a=0$ and its
minimum $0$ in the limit $\a \to \infty$.

\begin{figure}[ht]
\bea
\hspace{-84mm}
\frac{\partial D^{(\rho)}}{\partial n_e}|_{n_e=n_c^+}
\nonumber
\eea
\vspace{-15mm}
\begin{center}
\leavevmode
\epsfxsize=0.4\textwidth
\epsfbox{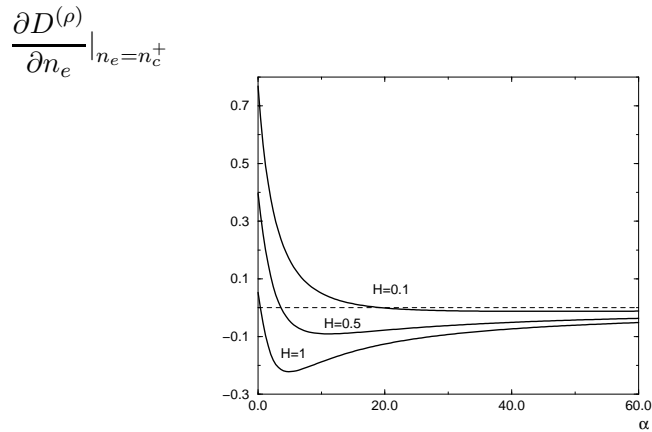}
\end{center}
\vspace{-1cm}
\caption{\label{fig:drs}
Derivative of the charge--stiffness with respect to the electron density 
at the crtitical electron density for a chain with $20$ percent impurities 
as a function of $\a$.}
\end{figure}

The derivative of the charge--stiffness with respect to the electron density
at $n_e=n_c^+$ changes sign as a function of $\a$ (see Fig.~\ref{fig:drs}).
\subsubsection{\sc Numerical Results}
\begin{figure}[ht]
\begin{center}
\leavevmode
\epsfxsize=0.4\textwidth
\epsfbox{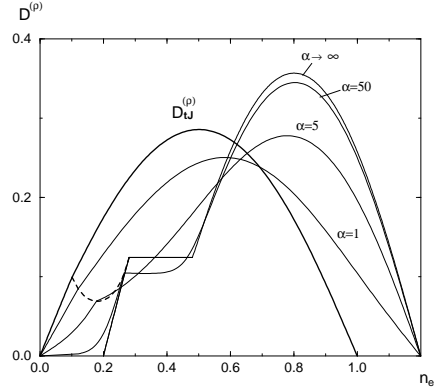}
\end{center}
\vspace{-1cm}
\caption{\label{fig:dc2}
Charge--stiffness for a chain with $20$ percent impurities and $H=0.1$
as a function of the electron density. The dashed line denotes the
stiffness at the critical electron density $n_c(\a)$.}
\end{figure}

\begin{figure}[ht]
\begin{center}
\leavevmode
\epsfxsize=0.4\textwidth
\epsfbox{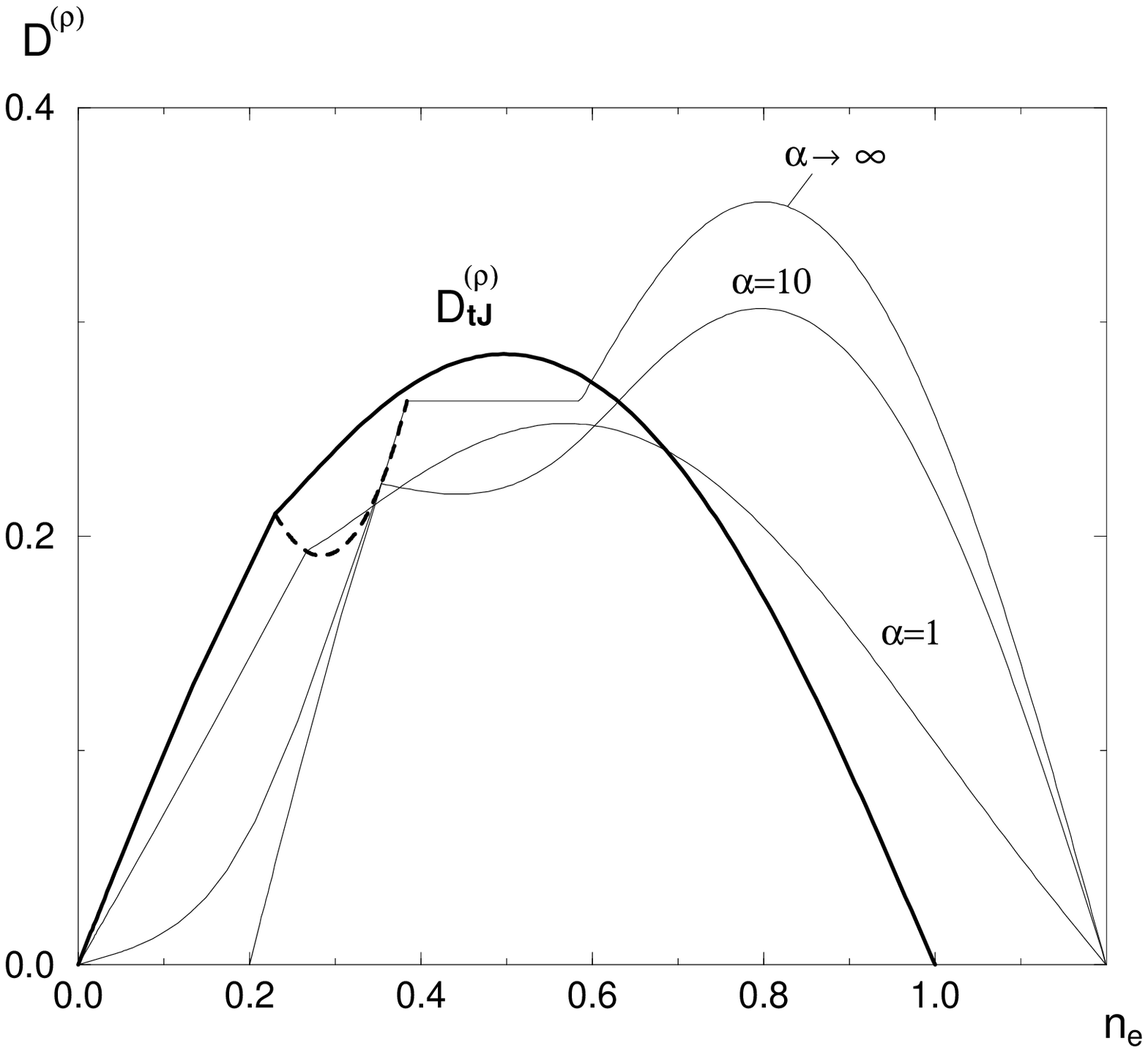}
\end{center}
\vspace{-1cm}
\caption{\label{fig:dc2hp5}
Charge--stiffness for a chain with $20$ percent impurities and
$H=0.5$ as a function of the electron density.}
\end{figure}

\begin{figure}[ht]
\begin{center}
\leavevmode
\epsfxsize=0.4\textwidth
\epsfbox{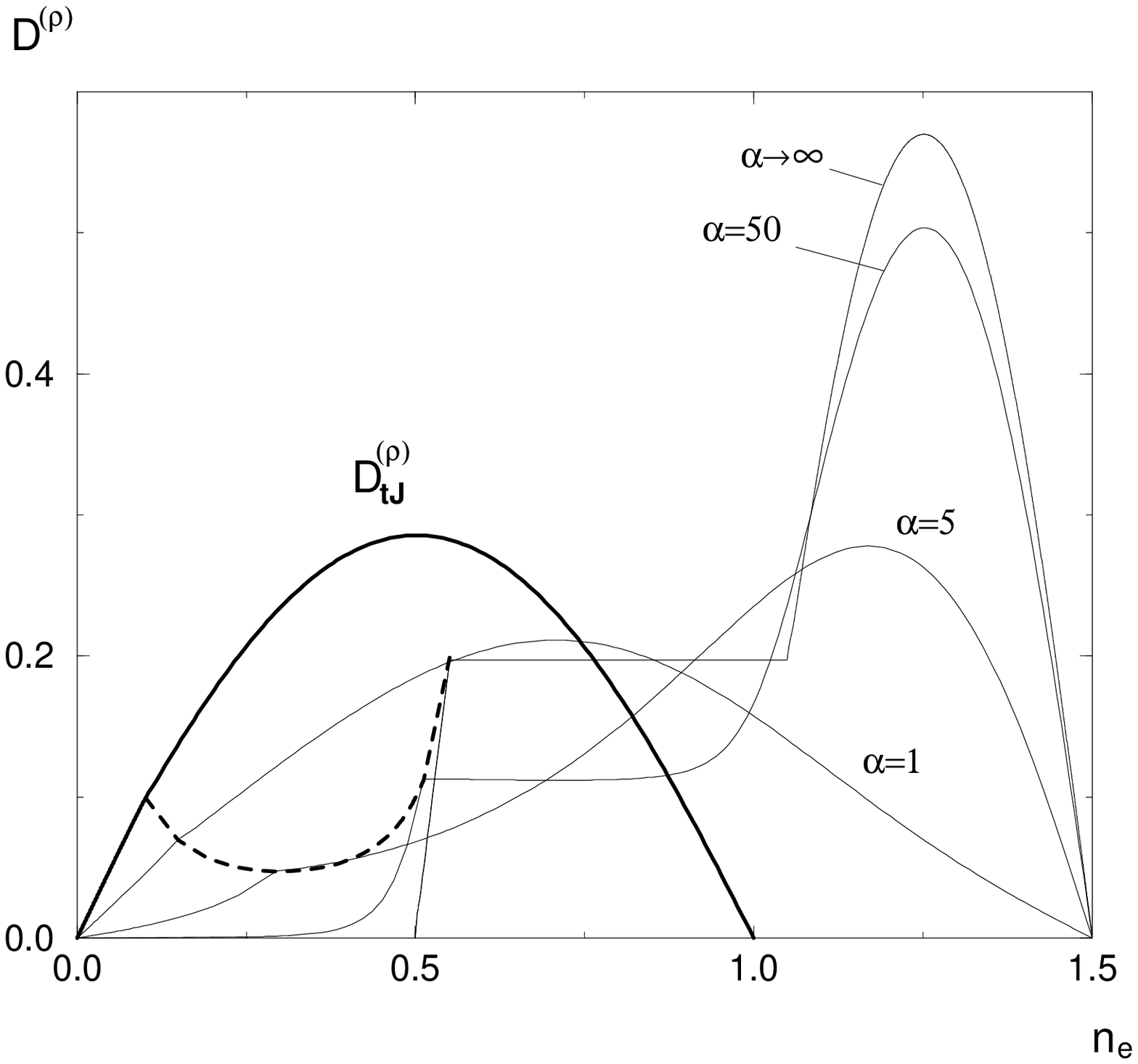}
\end{center}
\vspace{-1cm}
\caption{\label{fig:dcp5}
Charge--stiffness for a chain with $50$ percent impurities 
and $H=0.1$ as a function of the electron density.}
\end{figure}

The results for the charge stiffness in systems with impurity
densities $n_i=0.2$ and two different values of the bulk field $H$
are depicted in Fig.~\ref{fig:dc2} and Fig.~\ref{fig:dc2hp5}. The
charge stiffness for $n_i=0.5$ is shown in Fig.~\ref{fig:dcp5}.  
For comparison we plot the result for the charge--stiffness $D^\rho_{tJ}$
for the $t$-$J$ model without impurities as calculated in \cite{bares:92}.
We note that the maximal allowed band-filling is larger than one as
the impurity sites can be doubly occupied. We see that for small
band-fillings above the critical density the charge-stiffness is
reduced as compared to the pure $t$-$J$ case. For larger band fillings
the stiffness is found to increase in the presence of impurities. This
is easily understood: due to the constraint of single occupancy the
stiffness vanishes as we approach half-filling in the $t$-$J$
chain. The impurity sites can be doubly occupied, which gives the
electrons ``space to move'' and leads to an increase in the
stiffness. For large fillings the stiffness increases with increasing
$\a$ because (as can be deduced from the $a\rightarrow\infty$ limit)
the impurity sites become (on average) closer and closer to being
doubly occupied, which again makes it easier for the electrons to move
along the $t$-$J$ sites. 
Last but not least let us discuss the limiting case $\a=\infty$ at
impurity density $n_i$, {\sl i.e.} there are $Ln_i$ impurity sites and
$L(1-n_i)$ $t$-$J$ sites. For electron densities $n_e$ smaller than
$n_i$ the (spin-up) electrons (in the ground state) occupy only
impurity sites which do not interact with the $t$-$J$ sites. Thus the
stiffness is identically zero. For electron densities
$n_i<n_e<n_i+(1-n_i)n_c$ the saturated ferromagnetic ground
state on the $t$-$J$ sites is formed whereas all impurity sites are
singly occupied. The stiffness is completely determined by the $t$-$J$
sites. For densities in the interval
$n_i+(1-n_i)n_c<n_e<2n_i+(1-n_i)n_c$ the impurity sites become doubly
occupied and the stiffness does not change. For $n_e>2n_i+(1-n_i)n_c$
all impurity sites are doubly occupied and the stiffness follows 
(up to a rescaling by $\frac{1}{1-n_i}$)
the $t$-$J$ curve above the critical density $n_c$ 
(see (\ref{dgs}) and (\ref{dgr})).

\begin{figure}[ht]
\begin{center}
\leavevmode
\epsfxsize=0.4\textwidth
\epsfbox{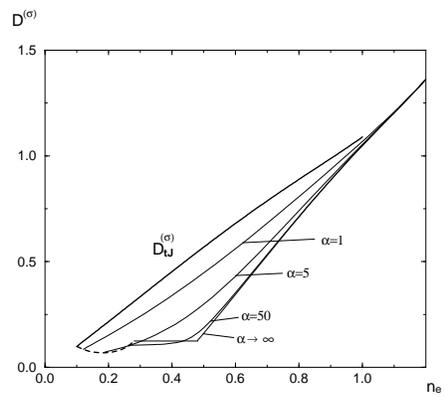}
\end{center}
\vspace{-1cm}
\caption{\label{fig:ds2}
Spin--stiffness for a chain with $20$ percent impurities 
as a function of the electron density.}
\end{figure}

The spin-stiffness for impurity density $0.2$ is shown in
Fig.~\ref{fig:ds2}. We see that the stiffness is decreased at low
fillings (this decrease is more pronounced for larger values of $\a$)
and approaches the ``pure'' $t$-$J$ value for large fillings. The
behaviour in the limit $\a \to \infty$ is the same as for the 
charge stiffness.
\section{\sc Impurity Phase-Shifts}
In this section we evaluate the phase shifts acquired by the
elementary excitations, holons and spinons, when scattering off the
impurity. The results give a good measure of the effects of the
impurity on excited states. In particular we can infer from the
phase-shifts how the impurity couples to spin and charge degrees of
freedom. We start by briefly reviewing some important facts about the
low-lying excitations in the $t$-$J$ model \cite{bares:91}. The
elementary excitations are collective modes of spin or charge degrees
of freedom. The spin excitations are called spinons and carry spin
$\pm \frac{1}{2}$ and zero electric charge. They are very similar to
the spin-waves in the Heisenberg XXX chain. The charge excitations are
called holons and antiholons, carry zero spin and charge $\mp e$. Thus
holons correspond to physical holes. At half-filling only holons can
be excited as the charge Fermi sea is completely empty.
The excitation energies are given by $\e_{c,s}$ defined in 
(\ref{dden2}). The respective physical momenta are given in terms of
the solutions of the following set of coupled integral equations
\bea
{\tt p}_s(\l)&=& -\theta(\l)-\int_{-A}^Ad\nu\ a_2(\l-\nu) {\tt p}_s(\nu)
+ \int_{-B}^B d\nu\ a_1(\l-\nu)\ {\tt p}_c(\nu)\ ,\nn 
{\tt p}_c(\l)&=& \int_{-A}^A d\nu\ a_1(\l-\nu)\ {\tt p}_s(\nu).
\label{mtm}
\eea
The mometum of {\sl e.g.} a holon-antiholon excitation is given by
$P_{c{\bar c}}={\tt p}_c(\La^p)-{\tt p}_c(\La^h)$ where $\La^p$ and
$\La^h$ are the spectral parameters of the holon and antiholon
respectively. We thus would define the physical holon momentum as
$p_c(\La^p) = {\tt p}_c(\La^p)-{\tt p}_c(B)$.
At half-filling the spinon ($p_s$) and holon ($p_c$) momenta are given
by 
\bea
p_s(\l)&=& 2\arctan(\exp(\pi\l))-\pi\ ,\nn
p_c(\l)&=& \frac{\pi}{2}+i\ \ln\left(
\frac{\Gamma(\frac{1-i\l}{2})}{\Gamma(\frac{1+i\l}{2})}
\frac{\Gamma(1+\frac{i\l}{2})}{\Gamma(1-\frac{i\l}{2})}\right).
\label{mtmhf}
\eea

The scattering matrix has been determined by means of Korepin's method
\cite{vladb,kor:79} in \cite{bares:91}. At half-filling the
spinon-spinon S-matrix $S(\l)$, the spinon-holon ($sc$) and
holon-holon ($cc$) scattering phases are given by
\bea
S(\l) &=& i\frac{\Gamma(\frac{1-i\l}{2})}{\Gamma(\frac{1+i\l}{2})}
\frac{\Gamma(1+\frac{i\l}{2})}{\Gamma(1-\frac{i\l}{2})}
\left(\frac{\l}{\l-i}I - \frac{i}{\l-i} P\right)\ ,\nn
\exp(i\varphi_{sc}(\l))&=& -i \frac{1+ie^{\pi\l}}{1-ie^{\pi\l}}\ ,\qquad
\exp(i\varphi_{cc}(\l))=
\frac{\Gamma(\frac{1+i\l}{2})}{\Gamma(\frac{1-i\l}{2})}
\frac{\Gamma(1-\frac{i\l}{2})}{\Gamma(1+\frac{i\l}{2})}
\label{smhf}
\eea
where $I$ and $P$ are the $4\times 4$ identity and permutation
matrices respectively. Below half-filling the S-matrices are given in
terms of the solution of integral equations.

The impurity phase-shifts can be computed by the standard method of
Korepin \cite{kor:79}, Andrei {\sl et. al.} \cite{al,ande:84}. 
In the most general case of a bulk magnetic field and arbitrary
filling factor the phase-shifts can be expressed only in terms of the
solution of a set of coupled integral equations, the analysis of which
is rather difficult. We therefore constrain ourselves to the case of a
microscopic number of holes in the half-filled ground state in the
absence of a bulk magnetic field.

The basic ingredient for computing impurity phase-shifts is the
quantization condition for factorized scattering of two particles with
rapidities $\l_{1,2}$ on a ring of length $N$ (including the impurity
site) 
\be
\exp(iNp(\l_1))S(\l_1-\l_2)e^{i\psi(\l_1)}=1\ ,
\label{qc}
\ee
where $p(\l)$ is the expression for the physical momentum of the
corresponding (infinite) periodic system, $S(\l)$ are the bulk
scattering matrices for scattering of particles $1$ and $2$, and
$\psi(\l_1)$ is the phase-shift acquired by particle $1$ when
scattering off the impurity. We note that the condition \r{qc}
incorporates the fact that there is no backscattering at the
impurity. For the present model the absence of backscattering follows
from the conservation laws for the rapidities: although momentum
is not a good quantum number for the ring with impurity, excited
states can still be characterized by the rapidity variables (see below
for an example). We would expect that if the impurity contained a
backscattering term mixing of states with different rapidities would
occur. This is not the case in the present model which indicates the
absence of backscattering. We note that the absence of backscattering
on the level of the ``bare'' Bethe Ansatz equations \r{eq:bagi} (which
describe the scattering of excitations over the empty ground state) is
not sufficient to deduce the absence of backscattering over the
antiferromagnetic ground state because the impurity gets dressed by
the holons and spinons in the ground state Fermi seas, and the
two-particle scattering processes between holons and spinons do
contain backscattering terms. Therefore the treatment of \cite{HPE}
does not apply in the present case.

In what follows we will extract the holon and spinon impurity
phase-shifts from the spinon-holon scattering state, for which the
condition \r{qc} turns into scalar equations for the scattering
phases, which after taking the logarithm read
\bea
Np_s(\l^h)+\varphi_{sc}(\l^h-\La^p)+\psi_s(\l^h)=0\ {\rm mod}\ 2\pi\ ,\nn
Np_c(\La^p)+\varphi_{sc}(\La^p-\l^h)+\psi_c(\La^p)=0\ {\rm mod}\ 2\pi\ .
\label{qchf2}
\eea
Here $\l^h$ and $\La^p$ are the rapidities of the spinon and holon
respectively. Comparing these conditions with certain quantities
(``counting functions'') that can be calculated from the Bethe Ansatz
solution one can then read off the boundary phase-shifts
$\psi_{s,c}$. 

Let us start by constructing the half-filled antiferromagnetic ground
state for even length $L$ of the host chain, where we furthermore
assume that $\frac{L}{2}$ is even as well. 
The ground state is obtained by choosing $M_1=\frac{L}{2}$,
$M^{(1)}=0$ in \r{baelog} and filling the half-odd integers $I^1_\b$
symmetrically around zero. In the thermodynamic limit this corresponds
to filling a Fermi sea of rapidities $\l^1_\b$ between $-\infty$ and
$\infty$, where the root density 
$\rho_s(\l^1_\b)=\lim_{L\to\infty}\frac{1}{(L+1)(\l^1_{\b+1}-\l^1_{\b})}$
is given in terms of the integral equation
\be
\rho_s(\l) = a_1(\l)\left(1-\frac{1}{L+1}\right) -\int_{-\infty}^\infty d\nu\
a_2(\l-\nu)\ \rho_s(\nu)\ .
\ee

The spinon-holon scattering state characterized by choosing
$M_1=\frac{L}{2}, M^{(1)}=1$ in the Bethe equations \r{baelog}. 
There are $\frac{L}{2}+1$ vacancies for the integers $I^1_\a$ and thus
one hole in the Fermi sea of $\l^1_\b$. 
We denote the rapidity corresponding to this hole by $\l^h$. The
rapidity corresponding to the holon is denoted by $\La^p$. 
The Bethe equations read 
\bea
\frac{2\pi}{L+1} I_\b &=& (1-\frac{1}{L+1}) \theta(\l_\b) -
\frac{1}{L+1}
\sum_{\b^\prime=1}^{\frac{L}{2}+1}\theta(\frac{\l_\b-\l_{\b^\prime}}{2})
+\frac{1}{L+1}\left[\theta(\frac{\l_\b-\l^h}{2})+
\theta(\l_\b-\La^p)\right],\nn
\frac{2\pi}{L+1} J &=& \frac{1}{L+1}\sum_{\b=1}^{\frac{L}{2}+1}
\theta(\La^p-\l_\b)+ \frac{1}{L+1}\theta(\frac{\La^p}{\a})
-\frac{1}{L+1}\theta(\La^p-\l^h),
\label{baehs}
\eea
where $J$ is a half-odd integer number.
In the limit $L\rightarrow\infty$ the distribution of roots $\l_\b$ is
described by a single integral equation for the density of roots
$\rho_s(\l)$, which is of Wiener-Hopf form but cannot be solved
in a form sufficiently explicit for the purpose of determining the
impurity phase-shifts. The main complication is that we need to take
into account all contributions of order $\frac{1}{L+1}$ and thus must
deal with the fact that the roots are distributed not between
$-\infty$ and $\infty$ but between two finite, $L$-dependent values
$-A$ and $A$. 
It can however be checked numerically that making the assumption that
the contributions due to the shift of integration boundaries will be of
higher order in $\frac{1}{L+1}$ as far as the impurity phase-shifts are
concerned (and thus taking $A=\infty$) yields the correct result.
The integral equation then can be solved by Fourier transform 
\be
\widetilde{\rho_s}(\omega) = \gt_0(\omega)+\frac{1}{L+1}\left\lbrace
\gt_1(\omega)e^{i\omega\l^h}-\gt_0(\omega)[1-2e^{i\omega\La^p}\right\rbrace ,
\label{dens} 
\ee
where $\widetilde{\rho_s}(\omega)$ is the Fourier transform of
$\rho_s(\omega)$ and where
$\gt_x(\omega)=\frac{\exp(-\frac{x}{2}|\omega|)}{2\cosh(\frac{\omega}{2})}$.

For the further analysis it is convenient to define counting
functions $z_s(\l)$ and $z_c(\l)$ 
\bea
z_s(\l) &=& \frac{L}{2\pi} \theta(\l) - \frac{1}{2\pi}
\sum_{\b=1}^{\frac{L}{2}+1}\theta(\frac{\l-\l_\b}{2})
+\frac{1}{2\pi}\left[\theta(\frac{\l-\l^h}{2})+\theta(\l-\La^p)\right],\nn
z_c(\La)&=& \frac{1}{2\pi}\sum_{\b=1}^{\frac{L}{2}+1}
\theta(\La-\l_\b)-\frac{1}{2\pi}\left[\theta(\La-\l^h)
+\theta(\frac{\La}{\a})\right].
\label{count}
\eea
Note that for any root {\sl e.g.} $\l_\a$ of \r{baehs} the counting
function takes the integer value $z_s(\l_\a)=I_\a$ by
construction. In the thermodynamic limit $\frac{1}{L+1}$ times the
derivative of $z_s$ yields the distribution function of rapidities
$\rho_s(\l)$. Straightforward 
integration of the density $\rho_s(\l)$ yields the following results
for the counting functions in the thermodynamic limit evaluated at the
rapidities of the spinon and holon respectively
\bea
2\pi z_s(\l^h)&=&(L+1)p_s(\l^h) + \varphi_{sc}(\l^h-\La^p)
+\phi_s(\l^h)=0\ {\rm mod}\ 2\pi\ ,\nn
-2\pi z_c(\La^p)&=&(L+1)p_c(\La^p) + \varphi_{sc}(\La^p-\l^h) + 
\phi_c(\La^p)=\pi\ {\rm mod}\ 2\pi\ ,
\label{cf}
\eea
where $p_{s,c}(\l)$ are the spinon/holon momenta \r{mtmhf},
$\varphi_{sc}(\l)$ is the bulk phase-shift for spinon-holon scattering
\r{smhf}, and
\be
\phi_s(\l)=-p_s(\l)\ ,\qquad
\phi_c(\l)=-p_c(\l)+\pi-\theta(\frac{\l}{\a})\ .
\ee
{From} these equations we can now infer the boundary phase shifts by
comparing them with the quantization condition \r{qchf2}, which yields
\bea 
e^{i\psi_s(\l)}&=& C\ \frac{\Gamma(\frac{1}{4}+\frac{i\l}{2})}
{\Gamma(\frac{1}{4}-\frac{i\l}{2})}
\frac{\Gamma(\frac{3}{4}-\frac{i\l}{2})}
{\Gamma(\frac{3}{4}+\frac{i\l}{2})}\ ,\nn
e^{i\psi_c(\l)}&=&-C^{-1}\ \frac{\a-2i\l}{\a+2i\l}\
\frac{\Gamma(\frac{1}{2}-\frac{i\l}{2})}
{\Gamma(\frac{1}{2}+\frac{i\l}{2})}
\frac{\Gamma(1+\frac{i\l}{2})}{\Gamma(1-\frac{i\l}{2})},
\eea
where $C$ is an overall constant factor of unit modulus that cannot be
determined within the Bethe Ansatz framework. 
Setting $C=-i$ we find that
\be
e^{i\psi_s(\l)}= e^{-ip_s(\l)}\ ,\qquad
e^{i\psi_c(\l)}= -\frac{\a-2i\l}{\a+2i\l}\ e^{-ip_c(\l)}\ ,
\ee
where $p_s$ and $p_c$ are the spinon and holon momenta at half-filling
respectively. 
This result is interpreted in the following way: for the
half-filled band doped with a finite number of holes the impurity site
essentially decouples from the host chain in the sense that spinons
and holons bypass it, which effectively shortens the lattice by one
site (see Fig. \ref{fig:hhalb})! \\ 
For the spinons this is the complete picture, whereas the holons still
acquire a phase shift due to the fact that the impurity site is
charged (recall that it is on average almost doubly occupied) and
therefore interacts with the holons passing it by. The holon
scattering phase has a pole at $\l=i\frac{\a}{2}$, which for $-2\leq\a
\leq-1$ lies on the physical sheet and therefore
corresponds to an impurity bound state. The restriction $\a<-1$ is
imposed in order to have a hermitean hamiltonian \cite{befr:95d}.
The impurity therefore has the interesting property to lead to a holon
bound state for sufficiently small negative $\a$ at half-filling.
\section{\sc Conclusion}
In this paper we have studied the effects of an integrable impurity in
a periodic $t$-$J$ model. The impurity couples to both spin and
charge degrees of freedom and the coupling strength $\a$ can be varied
continuously without losing integrability.
The two limiting cases $\a=0$ and $\a=\infty$ have been shown to be
described by effective (ground state) hamiltonians of a $t$-$J$ model
with one extra site, and a decoupled impurity in a $t$-$J$ chain with
twisted boundary conditions.

At zero temperature we have calculated the impurity magnetization and
particle number for arbitrary band filling and bulk magnetic field.
The impurity susceptibilities have been shown to exhibit the same
types of singularities as the corresponding bulk susceptibilities.
Similarly the low-temperature specific heat of impurity and bulk
have the same temperature dependence.
Transport properties have been determined through the calculation of
spin and charge stiffnesses and finally the impurity phase shifts have
been calculated for the half-filled band.

The supersymmetric $t$-$J$ model belongs to the class of Luttinger liquids
with repulsive electron-electron interactions. The effects of potential
impurities of the ``weak-link'' type were first investigated in
\cite{lupe:74,matt:74}.  It was found that the system flows to a
strong-coupling fixed point characterized by the vanishing of the dc
conductivity. The physics of the impurity studied here is quite different:
the dc-conductivity is unchanged by the presence of a single impurity.

As argued above the type of impurity considered here does not seem to
contain backscattering terms on the level of the dressed excitations
(holons and spinons). It would be interesting to verify this assertion
by explicitly constructing the continuum limit of the model. However,
due to the complicated structure of the impurity hamiltonian this is a
difficult undertaking. The argument given above suggest that it is
impossible to construct an impurity model containing backscattering
off the impurity by means of the Quantum Inverse Scattering Method
through the standard intertwining relation ``$RTT=TTR$'': the
rapidities of the elementary excitations will always be conserved
quantities and are not affected by the scattering off the impurity.
Clearly ``generic'' impurities ought to contain backscattering as only
special potentials are reflectionless. {}From that point of view the
integrable impurity considered in the present work is very special.
The situation is similar to the (multichannel) Kondo model (viewed as
a $1-d$ system).
One may speculate that like for the case of a Kondo-impurity in a
Luttinger liquid a backscattering term will drive the system to a new
fixed point \cite{frjo:96} so that the integrable impurity would
represent an unstable fixed point in the sense of the renormalisation
group.
This is known to be the case for the spin system of Andrei and
Johannesson \cite{soea:93}. 
As we have seen the integrable impurity nevertheless leads to
interesting physical consequences.

\vspace*{1cm}

\begin{center}
{\large \sc Acknowledgements}
\end{center}
We are grateful to A. Tsvelik and A.Jerez for important discussions
and suggestions. F.H.L.E. is supported by the EU under Human Capital
and Mobility fellowship grant ERBCHBGCT940709. He thanks the ITP at
Hannover, where part of this work was performed, for hospitality. This
work has been supported in part by the Deutsche Forschungsgemeinschaft
under Grant No.\ Fr~737/2--1.
\appendix
\section{\sc Transformation of the BAE}
To show the equivalence of the two sets of BAE
(\ref{eq:bagitwi}) and (\ref{eq:bagist}) we use a method due to
Woynarovich \cite{woyn:83} and Bares {\sl et al}\cite{bares:92}.
We express the second set of (\ref{eq:bagist}) as a polynomial of 
degree $M^{(1)}+N_\downarrow$
\be P(w)=\prod_{j=1}^{M^{(1)}+N_\downarrow-1}
(w-\l_j-{i\over 2})(w-i{\a \over 2})-
e^{-i\phi_c}\prod_{\b=1}^{M^{(1)}+N_\downarrow-1}(w-\l_j+{i \over
2})(w+i{\a \over 2})=0
\label{p1}
\ee 
and identify the first $M^{(1)}$ roots of (\ref{p1})
$w_1,\ldots,w_{M^{(1)}}$ with
$\l_1^{(1)},\ldots,\l_{M^{(1)}}^{(1)}$. Using the residue theorem we
obtain the following equality:  
\be \sum_{j=1}^{M^{(1)}}{1 \over
i}\ln \left[\frac{\l_l-\l_j^{(1)}+{i \over 2}}{\l_l-\l_j^{(1)}
-{i \over 2}}\right]
=\sum_{j=1}^{M^{(1)}}{1 \over 2\pi i} \oint_{{\cal C}_j}dz{1 \over i}\ln
\left[ \frac{\l_l-z+{i \over 2}}{\l_l-z-{i \over 2}}\right]
\frac{d}{dz} \ln(P(z)) 
\ee 
where ${\cal C}_j$ is a small contour
including $\l_j^{(1)}$. Deforming the contour and denoting the $N_\downarrow$
other roots of (\ref{p1}) with $w'_1,\ldots,w'_{N_\downarrow}$ we
arrive at the following equality
\be \sum_{j=1}^{M^{(1)}}{1 \over i}\ln
\left[\frac{\l_l-\l_j^{(1)}+{i \over 2}}{\l_l-\l_j^{(1)}-{i \over 2}}\right]=
-\sum_{j=1}^{N_\downarrow}{1 \over i}\ln \left[\frac{\l_l-w'_j+{i
\over 2}}{\l_l-w'_j-{i \over 2}}\right]+{1 \over i} \ln
\left[\frac{P(z_n)}{P(z_p)}\right],
\label{i1}
\ee where the last term comes from integration aorund the branch cut
extending from $z_n=\l_l+i/2$ to $z_p=\l_l-i/2$. Using the form of
$P(w)$ and substituting (\ref{i1}) into the first equation of
(\ref{eq:bagist}) we obtain the first equation of (\ref{eq:bagitwi})
with the according twist--angle
\be \left( \frac{\l_j-{i\over 2}}{\l_j+{i\over 2}} \right)^{L} \left(
\frac{\l_j-{\alpha+1 \over 2}i}{\l_j+{\alpha+1 \over 2}i} \right)=
e^{-i(\phi_c+\phi_s)} \prod^{N_\downarrow}_{\alpha=1} \frac
{\l_j-\lambda^{(1)}_\alpha-{i\over 2}}
{\l_j-\lambda^{(1)}_\alpha+{i\over 2}}\ , \quad j=1,\ldots,N_e \ ,
\ee 
The second equation of (\ref{eq:bagitwi}) can be obtained by the same
steps starting with the first equation of (\ref{eq:bagist}). The
twist--angles are related by $\phi_s=\phi_\uparrow-\phi_\downarrow$
and $\phi_c=\phi_\downarrow$ \cite{bares:92}.  
In \cite{esko:92,foka:93} it was shown that the BAE
(\ref{eq:bagist}) for the $t$-$J$ model without impurity can be
obtained by means of the QISM starting with a fermionic vacuum with
all spins up. The corresponding vacuum state of the impurity model is
given by a bosonic doubly occupied impurity site and all other sites
occupied with spin up electrons. The algebraic Bethe--Ansatz
starting from this vacuum can also be constructed.
\section{\sc The three site model}
The Bethe ansatz states do not form the complete set of eigenstates of
the system but are the highest-weight states of the $gl(2|1)$ superalgebra
(taking the Lai solutions of the BAE). Complementing the Bethe ansatz
states with those obtained by the action of the $gl(2|1)$ lowering
operators one obtains additional eigenstates. The completeness of this
extended Bethe ansatz has been proven for some models as the spin-${1
\over 2}$ Heisenberg chain, the supersymmetric $t$-$J$ model and the
Hubbard model \cite{foka:93,fata:84,eks:92}.
In this appendix we present a completeness analysis for impurity
system considered here on a chain with three sites. This nontrivial
example shows that the picture of \cite{foka:93,fata:84,eks:92} seems
to hold in the present model as well. A detailed analysis of the
general case is outside the scope of the present paper.

We need to consider the action (on states given by the Bethe Ansatz)
of the spin lowering operator $S^-=\sum_{i=1}^{L+1} X_i^{\downarrow
\uparrow} $ and the supersymmetry operators $Q_\sigma^\dagger$
($Q_\sigma$ in the Sutherland case), which are given by 
\be
Q_\downarrow = \sum_{i=1,i \neq 2}^{L+1} X_i^{0 \downarrow}
+\sqrt{\a+1}X_2^{0 \downarrow} +\sqrt{\a}X_2^{\uparrow 2}\ ,\quad
Q_\uparrow = \sum_{i=1,i \neq 2}^{L+1}X_i^{0 \uparrow}
+\sqrt{\a+1}X_2^{0 \uparrow} -\sqrt{\a}X_2^{\downarrow 2}\ .
\ee
They are seen to satisfy the commutation relations 
\be
\{Q_\uparrow,Q_\downarrow\}=0\ , \qquad Q_\sigma^2=0\ , \qquad 
[{\cal H},Q_\sigma]=0\ .
\ee
The BA states obtained by the Lai solution starting with empty
sites are characterized by $Q_\sigma |\Psi_{Lai}\rangle=0$. The
respective Sutherland solutions by $Q_\sigma^\dagger
|\Phi_{Suth.}\rangle=0$. Solving the BAE (\ref{eq:bagitwi}) and
(\ref{eq:bagist}) with vanishing twist angles for the simplest case
$L=2$ (recall that $L$ is the length of the host chain) and then
constructing the corresponding $gl(2|1)$ multiplet by acting with all
possible combinations of raising generators we obtain the following
complete set of eigenstates ($\lambda = {1 \over 2}\sqrt{\a+1 \over \a+3}$)
\vskip .3cm 

\begin{tabular}{|c|c|c|c|} \hline
Energy & $\sharp$ & BA Lai & BA Sutherland \\ \hline
0 & 4 & vacuum &\ $\l_1=-\l_2=\lambda \quad
	\l_1^{(1)}=-\l_2^{(1)}=\sqrt{\a \over \a+3}$\quad  \\
  &   & $|\Psi_0\rangle=|0\rangle$ & $ |\Phi_0\rangle =
Q_\uparrow^\dagger Q_\downarrow^\dagger |\Psi_0\rangle$ \\ 
\hline ${6+2\a \over \a+2}$ & 8 & $\l_1=\l \quad g=-{1 \over
2}\left[\sqrt{\a+1}+i\sqrt{\a+3}\right]$ & $\l_1=\l \quad
\l_1^{(1)}=\l {\a \over \a+1} $ \\ 
  &   & $|\Psi_1\rangle=g^*|\uparrow 0 0\rangle +g |0 0 \uparrow
\rangle +|0 \uparrow 0\rangle $ &  $ |\Phi_1\rangle=
Q_\uparrow^\dagger Q_\downarrow^\dagger |\Psi_1\rangle$ \\ \hline
${6+2\a \over \a+2}$ & 8 & $\l_1=-\l$ & $\l_1=-\l \quad 
\l_1^{(1)}=-\l {\a \over \a+1} $ \\
  &   & $|\Psi_2\rangle=g |\uparrow 0 0\rangle +g^* |0 0 \uparrow
\rangle +|0 \uparrow 0\rangle $  & $ |\Phi_2\rangle =
Q_\uparrow^\dagger Q_\downarrow^\dagger |\Psi_2\rangle$ \\ \hline
${12+4\a \over \a+2}$ & 12 & $\l_1=-\l_2=\l $ &  vacuum \\
  &   & $|\Psi_3\rangle=|\uparrow \uparrow
0\rangle-\sqrt{\a+1}|\uparrow 0 \uparrow\rangle + |0 \uparrow
\uparrow\rangle$ & $ |\Phi_3\rangle =Q_\uparrow^\dagger
Q_\downarrow^\dagger |\Psi_3\rangle= |\uparrow 2\uparrow \rangle $ \\
\hline
${4 \over \a+2}$ & 4 & $\l_1=-\l_2={i \over 2} \quad \l_1^{(1)}= 0 $ &
$\l_1=0$\\   &   & $|\Psi_4\rangle = Q_\uparrow Q_\downarrow
|\Phi_4\rangle $ & $|\Phi_4\rangle =|\uparrow 2 \downarrow\rangle 
-|\downarrow 2 \uparrow\rangle $ \\ \hline
\end{tabular} 

\vskip .3cm
The Sutherland solutions for the states $|\Phi_{0,1,2}\rangle$ and the
Lai solution for $|\Psi_4\rangle$ 
\footnote{The solution of the BAE exists but the norm of the corresponding
state vanishes.}  
do not exist in the case $\a=0$.  In this case the $Q_\sigma ^{(\dagger)}$
commute with $X_2^{22}$ and the states decompose into the 27 states
corresponding to the Lai-states without double occupancy and the 9 states
of the Sutherland solution with doubly occupied impurity site
\footnote{The multiplet of dimension 12 splits in one 7 and one 5
dimensional multiplet as $Q_\downarrow^\dagger (\a=0) |\uparrow
\uparrow \uparrow\rangle =0$}. This is in agreement with the form of the
hamiltonian in the limit $\a \to 0$ given in (\ref{hgl}). 

Let us conclude this appendix with some simple considerations
concerning the question of whether the impurity contributions to
particle number and magnetization are indeed concentrated at the
impurity. 
The lowest energy state with one electron on a lattice of arbitrary
length is given by $Q_\sigma^\dagger |0\rangle$. 
Using this fact we are able to directly compute the electron density
at the impurity for this state and we find that
$\langle n_2 \rangle=\frac{1+\a}{L+1+\a}$. In order to compare this
with (\ref{idifer}) we need to take into account that (\ref{idifer})
is obtained in the thermodynamic limit, {\sl i.e.} we need to take
$L\gg \a+1$. We then find that $N_\a=\frac{1+\a}{L}$, which is in
agreement with $\langle n_2 \rangle$.

Similarly the lowest energy state above the critical density $n_c$ is
given by $Q_\downarrow^\dagger|\Psi_{Ferr.}\rangle$. By the action
of the $Q$-operator a spin--down electron is generated with probability
$(\a+1) \times (1-N_\a)$ and a doubly occupied impurity--site with
probability $\a \times N_\a$. Taking into account the
normalization--factor ${1 \over L(1-n_e)}$ we obtain the following
result
\be
{\partial N_\a \over \partial n_e}|_{n_e=n_c^+}=
{\Delta N_\a \over \Delta n_e}=L \Delta N_\a=
{\a+1-N_\a \over 1-n_e}+o({1 \over L}).
\ee
This coincides with (\ref{abnc}). {}From these simple examples we deduce
that the assumption that the impurity contributions to magnetization
and particle number are located at the impurity is a very reasonable
one. 
\section{\sc $gl(2|1)$ Invariance of the Model}
In this appendix we show by explicit computation that the model
\r{hamil} is $gl(2|1)$-invariant. We start by expanding $R$-matrix and
L-operators around infinite spectral parameters
\bea
R_{33}(\l)&\sim& \Pi^{00} +\frac{i}{\l}\left(I-\Pi\right)^{00}\ ,\nn
{\cal L}_{33}^n(\l)&\sim& I^{0n}+\frac{i}{\l}\left(\Pi-I\right)^{0n}\ ,\nn
{\cal L}_{34}(\l)&\sim& I^{02}+\frac{i}{\l}\left({\cal
L}-(2+\a)I\right)^{02}\ , 
\label{c1}
\eea
where we denoted the auxiliary space by $0$, $n$ labels the quantum
spaces over the $t$-$J$-like sites, and the impurity sits at site $2$.
This leads to the following expansion of the monodromy matrix
\be
T(\l)\sim I+\frac{i}{\l}\left[\sum_n (\Pi-I)^{0n}+({\cal
L}-(2+\a)I)^{02}\right]=:I+\frac{i}{\l}Y\ .
\label{c2}
\ee
Inserting \r{c1} and \r{c2} into the intertwining relation
\be
R_{33}(\l-\mu)\left(T(\l)\otimes T(\mu)\right)=
\left(T(\mu)\otimes T(\l)\right)R_{33}(\l-\mu)
\ee
we obtain the following equations
\bea
&&(-1)^{\eps_a\eps_{\ap}}\left[Y^{b\ap}-\delta_{b\ap}\right]
T(\mu)^{a\bp}+\delta_{a\ap} T(\mu)^{b\bp}=\nn
&&\qquad(-1)^{\eps_\ap\eps_\bp+\eps_b(\eps_\bp+\eps_a)} T(\mu)^{a\bp}
\left[Y^{b\ap}-\delta_{b\ap}\right]+(-1)^{(\eps_a+\eps_\ap)\eps_b} 
\delta_{b\bp} T(\mu)^{a\ap}\ .
\label{c3}
\eea
Setting $a=\bp$, multiplying \r{c3} by
$(-1)^{\eps_a(\eps_a+\eps_\ap)}$ and then summing over $a$ we arrive
at
\be
[Y^{b\ap},\tau(\mu)] = 0\ ,
\ee
where $\tau(\mu)$ is the transfer matrix of the system. Dropping some
constants we therefore find that
\be
[Q_{ab},\tau(\mu)] = 0\ ,\quad Q_{ab} = \sum_n X_n^{ab}+{\cal L}^{ab}\
,\quad a,b=1\dots 3,
\ee
where we use the correspondences $1\leftrightarrow\up$,
$2\leftrightarrow\da$, $3\leftrightarrow0$ for $X^{ab}$. The
operators $Q_{ab}$ form a complete set of generators for $gl(2|1)$,
which establishes the invariance of our model.

\newpage
\centerline{\large \bf Figure Captions}

\end{document}